\documentclass[12pt] {article}
\usepackage{amstex}
\usepackage{epsfig}

\newcommand{\nc}{\newcommand}
\nc{\beq}{\begin{equation}}
\nc{\eeq}{\end{equation}}
\nc{\beqa}{\begin{eqnarray}}
\nc{\eeqa}{\end{eqnarray}}
\newcommand{\bea}{\begin{eqnarray}}
\newcommand{\eea}{\end{eqnarray}}

\newcommand{\bean}{\begin{eqnarray*}}
\newcommand{\eean}{\end{eqnarray*}}
\newcommand{\ecran}[2]{ $\left( \hspace{-0.2cm} { \raisebox{-0.1cm} {~\shortstack{ $#1$ \\ #2}} } \right)$ }
\def\agt{
\mathrel{\raise.3ex\hbox{$>$}\mkern-14mu\lower0.6ex\hbox{$\sim$}}
}
\def\alt{
\mathrel{\raise.3ex\hbox{$<$}\mkern-14mu\lower0.6ex\hbox{$\sim$}}
}

\begin {document}
\bibliographystyle{unsrt}    

\hspace{12cm}PM$-99$/54

\vspace{2cm}

\Large
{\centerline{\bf Dual generalizations of sine-Gordon field theory}}
\vspace{3mm}
{\centerline{\bf and integrability submanifolds in parameter space}
\vspace{3mm}
{\centerline{\bf }
\vspace{2mm}

\large

\vspace{1cm}

\centerline {\bf P. Baseilhac$^{1,2,3}$ and D. Reynaud$^{3}$}

\small\normalsize
\vspace{5mm}

\centerline {$^{1}$\it  School of Physics, Korea Institute for Advanced Study,}

\vspace{1mm}

\centerline {\it Seoul, 130-012, Korea}
\vspace{7mm}

\centerline {$^{2}$\it Department of Mathematics, University of York,}

\vspace{1mm}

\centerline {\it Heslington, York YO10 5DD, UK}
\vspace{7mm}

\centerline{$^{3}$ \it Laboratoire de Physique Math\'ematique et Th\'eorique} 

\vspace{1mm}

\centerline{\it UMR CNRS 5825, Universit\'e Montpellier II}

\vspace{1mm}

\centerline {\it Place E.~Bataillon, 34095 Montpellier, France}

\vspace{1cm}

\begin{abstract}  
The dual relationship between two $n-1$ parameter families of quantum field 
theories based on extended complex numbers is investigated in two dimensions. 
The non-local conserved charges approach is used. The lowest rank affine Toda field
theories are generated and identified as integrability submanifolds in parameter space. A
truncation of the model leads to a conformal field theory in extended
complex space. Depending on the projection over usual complex space
 chosen, a parametrized central charge is calculated.  
\end{abstract}

\vspace{1mm} 
\newpage
\setcounter{page}{2}
\section{Introduction}\label{Introduction}
In statistical mechanics and quantum field theory (QFT), duality plays an
important role for exploring strong coupling regime from knowledge of the weak
coupling behavior.  The
sine-Gordon and massive Thirring models are famous examples \cite{1}.
In this context, affine Toda field theories (ATFTs) are one-parameter families
of quantum integrable massive field theories possessing such duality properties \cite{Ari}.

Recently, we introduced \cite{multisine,4} and studied a new $n-1$ parameter 
$(\{\alpha_a\},\{\beta_a\})$ family of quantum field theories, called 
multisine-Gordon (MSG) models. The general construction
of the MSG in terms of the extended trigonometry associated to the extended 
complex numbers \cite{5}, namely multicomplex (MC) numbers, was obtained starting
 from a generator \  $e$ \ such that\ \ $e^n=-1$. For 
$(\{\alpha_a\}\in {\mathbb R},\{\beta_a\} \in{\mathbb R})$, we have shown 
that these MSG models provide a 
unifying representation for a wide variety of integrable QFTs possessing dual 
representations. In particular, integrable models studied a few years ago \cite{6},
like integrable deformations of non-linear $\sigma$ models or massive Thirring 
coupled with ATFTs were recovered subject to restrictions on parameter space.

However, a breakthrough comes in \cite{Pascal} from the writing of the 
extended trigonometric functions (multisine functions), which appeared in the
MSG potentials, in terms of the natural multicomplex extension of vertex 
operators, namely MC-vertex operators.  Using this framework, we investigated
the existence of non-local conserved charges. In lower-dimensional QFTs, it is well-known that non-local
conserved charges may appear, which generate symmetries characterized by braiding
relations \cite{7}. They provide a powerful tool for studying non-perturbative
 effects \cite{8}. A set of equations associated with the conservation of 
these non-local charges and their algebraic structure was thus obtained.

In this paper, we investigate the dual relationship between two Lagrangian
representations generalizing the sine-Gordon model in the previous meaning.
Up to restrictions on the parameter space, we show that lowest-rank ATFTs
are generated by MC-algebras. Next, description of various integrable perturbations of 
 conformal field theories as different projections over the usual complex space of the same MC-vertex operator is studied.

In section 2, we briefly recall and reinterpret conveniently some results
 obtained in \cite{multisine,4,Pascal} which
constitute the basic ingredient of the following sections. For  
$(\{\alpha_a\}\in{\mathbb R},\{\beta_a\} \in i{\mathbb R})$ (or $(\{\alpha_a\}\in{i\mathbb R},
\{\beta_a\}\in{\mathbb R})$, a generic solution
 of the equations associated to non-local currents conservation is given. Due to their structure, it naturally emerges 
 a dual relation between the parameters of the model (which appear in the
 potential through MC-vertex operators) and those involved in the 
conserved currents.  To first order in conformal perturbation theory (CPT), it is then
 possible to introduce a ``dual'' family of Lagrangian representations :  the ``dual'' potential is built using the 
expression in MC-space of conserved currents associated to the original one.

In Section 3, imposing a quantum algebraic structure to the non-local conserved
 charges, we solve the previous equations.  Each solution is 
associated with a multicomplex space of dimension $n$, a specific MC-algebra
 and different ratios of the parameters. For 
 $(\{\alpha_a\}\in{\mathbb R},\{\beta_a\} \in i{\mathbb R})$, the underlying hidden
symmetry of MSG model in each case is identified to a quantum universal
enveloping algebra (QUEA) based on an affine Lie algebra ${\hat{\cal{G}}}$.
This approach provides a unifying ``parametrized'' description of 
lowest-rank QUEAs based on $A_r^{(1)}$ for $r\leq 3$, $A_{2r}^{(2)}$ for $r\leq 2$,
 $D_4^{(1)}$, $(B_r^{(1)},A_{2r-1}^{(2)})$ for $r\leq 3$, $(C_2^{(1)},D_{3}^{(2)})$
and $(G_2^{(1)},D_{4}^{(3)})$.

We show in section 4 how ATFTs are related to MSG models. In particular, simple
 relations between the multicomplex dimension $n$ and the rank of
 $\hat{\cal{G}}$, the kind of MC-algebra (characterized by $m_a$) and
the Kac labels (denoted $n_a$) are obtained. For $n=3$ and $n=4$, we describe the 
(dual-)MC-algebras generating $A_2^{(1)}$, $C_2^{(1)}$, $D_3^{(2)}$ and $A_3^{(1)}$
ATFTs. 

In section 5, we show that whereas the multisine-potential do not depend on the
projections of MC-algebras over the usual complex space, its understanding in terms of perturbed
conformal field theory (CFT) does. We identify two kinds of perturbed CFT 
through the introduction of a (real) multicomplex charge at infinity. For each 
projection, a ``parametrized'' central charge is computed.

Some conclusions and perspectives are drawn in section 6.

\section{Dual conserved currents in extended sine-Gordon}
In  \cite{multisine,4}, we introduced the natural extension of the 
sine-Gordon field theory in the $n$-dimensional multicomplex space. This model,
generated by the fundamental multicomplex number\,\footnote{More precisely, $e$
is denoted $e_{(n|m)}$ in ref. \cite{4} due to its eventual substructure.} $e$
 \cite{mc1,mc2},
such that $e^n=-1$, describes a family of $n-1$ parameter
quantum field theories with $n-1$ scalar fields which interact through a 
multisine potential \cite{multisine,4}.
Its Euclidian action can generally be expressed in terms of the extension in 
MC-space of standard vertex operators \cite{Pascal} :   
\beqa
{\cal{A}}^{(n|m)}(\eta)= \frac{1}{4\pi}\int d^2z \partial_z\Phi\partial_{\overline z}\Phi + \frac{\lambda}{n\pi}\int d^2z \Big(x^{(0)}+...+ x^{(n-1)}\Big)\label{action}  
\eeqa
with :
\beqa
x^{(l)}= \exp\big(\eta^{(l)}.\Phi(z,\overline{z})\big),\label{xl}
\eeqa
where we define :
\beqa
m = 2\sum_{a=0}^{\frac{n}{2}-1}m_{a} \ \ \mbox{for $n$ even\ \ \ and\ \ \ } 
m = 2\sum_{a=0}^{\frac{n-1}{2}-1}m_{a}+m_{(n-1)/2} \ \ \mbox{for $n$ odd}\label{m}
\eeqa 
characterize the kind of MC-algebra and $\Phi(z,\overline{z})$ is the fundamental
 ($n-1$-components)-field of the theory. Note that the index $l$ denotes 
 the $l$-{\it th}\ multicomplex conjugation of any MC-number. In \cite{multisine}, we considered only unimodular MC-numbers. The unimodularity condition was implemented on MC-vertex operators (\ref{xl}) through : 
\beqa
\label{unimod}
||x||^{n}=\prod_{l=0}^{n-1}x^{(l)}=1,
\eeqa
where $||\cdot ||$ is the pseudo-norm associated to MC-algebra
 (see refs. \cite{mc1,mc2} for details). Depending on the value of $n$ (even or odd cases),
  expression (\ref{xl}) differs. For $n$ even, the MC-vertex operator (\ref{xl})
   is defined as follow. Firstly, we introduce the bilinear relation \cite{Pascal}:
\beqa
. \ \ \ \ \ : \big({\mathbb M \mathbb C}_{(n|m)}\big)^{n-1} \times \big({\mathbb M \mathbb C}_{(n|m)}\big)^{n-1} \longrightarrow {\mathbb M \mathbb C}_{(n|m)}
\eeqa
which can be understood as an extension of the standard scalar product in the
multicomplex valued vector space. MC-vertex operators depend on parameters 
$(\{\alpha_a\},\{\beta_a\})$ through the relation :
\beqa
{\eta}^{(l)}&=& \big[ \alpha_0 P_{l},...,\alpha_{\frac{n}{2}-1} P_{\frac{n}{2}-1+l}; \beta_0 Q_{0;l},...,\beta_{\frac{n}{2}-2} Q_{\frac{n}{2}-2;l} \big] \in \big({\mathbb M \mathbb C}_{(n|m)}\big)^{n-1},\label{etak}
\eeqa
where the parameters $(\{\alpha_a\},\{\beta_a\})\in {\mathbb{C}}$ and 
$\{P_{a+l},Q_{a;l}\}$ generate the MC-algebra\,\footnote{For $n$ even
and ``mimimal'' representations (i.e $m_a=1$ for all $a$), the generators $P_a$
are expressed \cite{multisine,Pascal} in terms of the fundamental multicomplex element $e$ as :
$P_a=\frac{2}{n}\sum_{j=0}^{n/2-1}\sin[(2a+1)j\frac{\pi}{n}]e^j$.} \cite{Pascal}. Consequently, if $\phi(z)$ denotes
the holomorphic part of the fundamental field $\Phi(z,{\overline z})$, 
e.g $\Phi(z,{\overline z}) =\phi(z)+{\overline {\phi(z)}}$, expression 
(\ref{xl}) is defined (since $\mathbb{C} \subset {\mathbb M \mathbb C}_{(n|m)}$) with :
\beqa
\eta^{(l)}.\phi(z)= \sum_{a=0}^{\frac{n}{2}-1} \Big[ \alpha_a P_{a+l}\phi_a(z)\Big] + \sum_{a=0}^{\frac{n}{2}-2} \Big[\beta_a Q_{a;l}\varphi_a(z) \Big]\label{d2}
\eeqa
and :
\beqa
&&Q_{a;l}=-P_{a+l}^2+\frac{m_a}{m_{n/2-1}}P^2_{n/2-1+l }, \ \ \ Q_{a;l}= Q_{a;l+n/2}= Q_{a +n/2;l}= Q_{a+n;l} = Q_{a;l+n},\nonumber
\eeqa
for\  $a\in \{ 0,...,n/2-2 \}$,\ \ \ \ $l\in \{ 0,...,n/2-1\} $\  \ \ with the conventions\  \ \ $\alpha_{a+n/2}=\alpha_a,\  \beta_{a+n/2}=\beta_a$\ \  and\ \  $m_{a+n/2}=m_a$.\\

In \cite{multisine, 4}, the MSG models were studied for {\it real} parameters.
Let us here focus on the parameter space restricted to :
\beqa
&&\alpha_a\in\mathbb{R} \ \ \ \ \hbox{and} \ \ \ \ \beta_a\in i\mathbb{R},\\
\hbox{or} &&\alpha_a\in i\mathbb{R} \ \ \ \ \hbox{and} \ \ \ \ \beta_a\in \mathbb{R},\nonumber
\eeqa
for all $a$. In these cases, we introduce the dual multicomplex element 
$\eta^{(k) \vee} = 2 \eta^{(k)} (\eta^{(k)}.\eta^{(k)})^{-1}$ 
(supposing that $(\eta^{(k)}.\eta^{(k)})$ is invertible which will be always 
satisfied in the sequel). Using MC-algebra \cite{Pascal} and equation (\ref{etak}), it leads
to the dual parameter space :
\beqa
&&\alpha_a^{\vee} = \frac{-2\alpha_a}{({\alpha_a}^2-\beta_a^2)}\ \ \ \  \  \mbox{for}\ \ a \in \{ 0,...,n/2-2\} ,\nonumber \\
&&\beta_a^{\vee} = \frac{-2\beta_a}{(\alpha_a^2-\beta_a^2)}\ \ \ \  \  \mbox{for}\ \  a \in \{ 0,...,n/2-2\} , \label{change} \\
&&\alpha_{n/2-1}^{\vee} = \frac{-2\alpha_{n/2-1}}{(\alpha_{n/2-1}^2-\sum_{a=0}^{n/2-2}\frac{m_a^2}{m_{\frac{n}{2}-1}^2} \beta_a^2)},\nonumber
\eeqa
and the dual MC-algebra generated by $e^\vee$ with :
\beqa
&& \frac{m_a^{\vee}}{m_{\frac{n}{2}-1}^{\vee}} = \frac{m_a}{m_{\frac{n}{2}-1}}\frac{(\alpha_a^2-\beta_a^2)}{(\alpha_{n/2-1}^2-\sum_{a=0}^{n/2-2}\frac{m_a^2}{m_{\frac{n}{2}-1}^2}\beta_a^2)}\ \ \ \  \  \mbox{for}\ \ a \in \{ 0,...,n/2-2\}\label{change2}
\eeqa
In the following, it is then more convenient to rewrite the action of the MSG (\ref{action}) in MC-space as :
\beqa
{\cal{A}}^{(n|m)}(\eta)&=& \frac{1}{4\pi}\int d^2z \partial_z\Phi\partial_{\overline z}\Phi + \frac{\lambda}{2\pi}\int d^2z \Phi_{pert}(\eta).\label{action2}
\eeqa
By analogy,
we can now consider the MSG model with action 
${\cal A}^{(n | m^{\vee})}(\eta^{\vee})$, where 
$m^{\vee}$ is defined as in eq. (\ref{m}) with eqs. (\ref{change2}). The perturbing operator for each model reads
respectively  :
\beqa
\Phi_{pert}(\eta) = \frac{2}{n} \sum_{l} J_{\eta^{(l)}} \overline{J}_{\eta^{(l)}}
\ \ \ \ \hbox{and} \ \ \ \ \Phi_{pert}(\eta^\vee) = \frac{2}{n} \sum_{l} J_{\eta^{(l)\vee}} \overline{J}_{\eta^{(l)\vee}}.\label{ver}
\eeqa
Considering now action (\ref{action}) as a perturbed $(n-1)$ free field CFT 
and following Zamolodchikov approach \cite{Zam}, it is possible to construct
$2n$ non-local conserved charges to first order in conformal perturbation 
theory (CPT). Let us consider the holomorphic MC-vertex operator 
$J^{(k)}=J_{{\eta'}^{(k)}}=e^{{\eta'}^{(k)}.\phi(z)}$ (and respectively, for
the antiholomorphic part, 
${\overline J}^{(k)}={\overline J}_{{\eta'}^{(k)}}=e^{{\eta'}^{(k)}.{\overline {\phi(z)}}}$).
Let us also suppose that its OPE with the $k^{th}$-perturbing term of the potential leads to a
derivative term : ${\overline\partial}J^{(k)} = \partial H^{(k)}$ and similarly
for the antiholomorphic part (whereas OPEs with any other $l\neq k^{th}$-perturbing
term yields to regular terms), e.g. the OPE reads :
\beqa
J_{{\eta'}^{(k)}}(z)x^{(l)}(w) \sim (z-w)^{-C^{k,l}({\eta'},\eta)} e^{({\eta'}^{(k)}+{\eta}^{(l)}).\phi(w)} +...\label{op}
\eeqa
with ${\eta'}^{(k)}$ defined as in (\ref{etak}) but with 
$\alpha_a\rightarrow\alpha'_a$, $\beta_a\rightarrow\beta'_a$ and
$m_a\rightarrow m'_a$. Since MC-exponents $C^{k,l}({\eta'},\eta)={\eta'}^{(k)}.{\eta}^{(l)}$ (reported in \cite{Pascal}) are of the form \ \  $C^{k,l}(\eta',\eta) = \sum_{a=0}^{\frac{n}{2}-1} C_a^{k,l} (\eta',\eta) (-P_a^2)$ in terms of MC-algebra, where\  $C_a^{k,l}({\eta'},\eta)$\  depend on $\alpha_a$, $\beta_a$, $\alpha_a'$, $\beta_a'$, then using MC-algebra, one easily shows that :
\beqa
J_{\eta^{'(k)}}(z) x^{(l)}(w) \sim \Big[ \sum_{a=0}^{\frac{n}{2}-1} (z-w)^{-C_a^{k,l}({\eta'},\eta)} (-P_a^2) \Big] e^{({\eta'}^{(k)}+{\eta}^{(l)}).\phi(w)} + \dots
\eeqa
It clearly implies that the conservation condition of the holomorphic current
$J^{(k)}$ does not depend on the choice of any multicomplex representation.
It remains to solve :
\beqa
C_a^{k,k}({\eta'},\eta)=2\ , \ \ \ \ C_a^{k,l}({\eta'},\eta)\in \mathbb{R_-}\ \ \ \hbox{for} \ \  k \neq l, \ \ \ \hbox{for all}\ a,\label{A}
\eeqa
with $(k,l)\in\{0,...,n/2-1\}$. Whereas this system of constraints looks 
overdeterminate, invariance under multicomplex conjugation of action 
(\ref{action}) leads to redundancies. In parameter space, the constraints (\ref{A})
reads :
\beqa
&&\Big(-\alpha'_a \alpha_a + \beta'_a \beta_a\Big) = 2 ,\ \ \ \ \ \ \Big(-\alpha'_ {\frac{n}{2}-1}\alpha_{\frac{n}{2}-1} + \sum_{a=0}^{\frac{n}{2}-2}\beta'_a \beta_a \frac{m'_a m_a}{m'_{\frac{n}{2}-1}m_{\frac{n}{2}-1}}\Big) = 2, \label{cont3} \\
&& \Big(\alpha'_a \alpha_a + \beta'_a \beta_a\Big) \in {\mathbb R}_{-} ,\ \ \ \  \ \ \Big(\alpha'_ {\frac{n}{2}-1}\alpha_{\frac{n}{2}-1} + \sum_{a=0}^{\frac{n}{2}-2}\beta'_a \beta_a \frac{m'_a m_a}{m'_{\frac{n}{2}-1} m_{\frac{n}{2}-1}}\Big) \in {\mathbb R}_{-}, \label{cont4}\\
&&\frac{m_a}{m_{\frac{n}{2}-1}}\beta'_a \beta_a  \in {\mathbb R}_{+}^{*} \ \ \ \ \ \mbox{and}\ \ \ \ \frac{m'_a}{m'_{\frac{n}{2}-1}}\beta'_a \beta_a  \in {\mathbb R}_{+}^{*},\label{contr5}
\eeqa
for all $a\in \{ 0,...,n/2-2 \}$. If eqs. (\ref{cont3}), (\ref{cont4}) and (\ref{contr5})
 are satisfied, the non-local currents $J^{(k)}$ are conserved to first order 
in CPT for all $k$ (and similarly for the antiholomorphic part). They generate
$2n$ non-local conserved charges :
\beqa
Q^{(k)} &=& \frac{1}{2i\pi} \Big( \oint_z dz J^{(k)}\ + \ \oint_{\overline z} d{\overline z} H^{(k)}\Big), \nonumber \\
{\overline Q}^{(k)} &=& \frac{1}{2i\pi} \Big(\oint_{\overline z} d{\overline z} {\overline J}^{(k)} \ +\ \oint_z dz {\overline H}^{(k)}\Big)\ \ \ \ \mbox{for}\ \ k\in[0,...,n-1]\label{mucur}.
\eeqa 
The non-locality is due to the fact that (anti-)chiral components $\phi$ and 
$\overline{\phi}$ of the MSG model are non-local with respect to the fundamental field $\Phi$.
From this property, braiding relations may arised between these charges.
It is now obvious to see that the parameters\ \  $\alpha_a' = \alpha_a^{\vee}$,\  $\beta_a' = \beta_a^{\vee}$
 and \ \ $m'_a/m'_{n/2-1}=m^{\vee}_a/m^\vee_{n/2-1}$\ \ satisfy 
 eqs. (\ref{cont3}), whereas eqs. (\ref{cont4}), (\ref{contr5}) still
constraint the parameter space $(\{\alpha_a\},\{\beta_a\})$ of MSG (\ref{action}).
Under these additional conditions on parameter space, conservation of
$J_{{\eta'}^{(k)}}=J_{\eta^{(k)\vee}}$ in the model associated to action ${\cal A}^{(n|m)}(\eta)$ is ensured. 
 It possesses $2n$ non-local conserved currents
$\{J_{\eta^{(k)\vee}},{{\overline J}_{\eta^{(k)\vee}}}\}$.  Consequently, since $C^{k,l}( \mbox{\small{(}} \eta^{\vee} 
\mbox{\small{)}}^{\vee} , \eta^{\vee} ) = C^{l,k}(\eta^{\vee} ,\eta )$,
 the whole set of 
non-local currents $\{J_{\eta^{(k)}},{{\overline J}_{\eta^{(k)}}}\}$ 
are similarly conserved to first order in CPT, in the model 
${\cal A}^{(n|m^{\vee})}(\eta^{\vee})$. It allows to define a duality relation
 between these two models, at least to order ${\cal O}(\lambda)$.
Note that the transformations (\ref{change}), (\ref{change2}) involving $n-1$ parameters, resulting of the 
current conservation, is completly independent of any particular structure
of non-local conserved charge algebra.

However, to define a consistent QFT, we have to consider carefully 
the renormalization group flows in such models. The crucial point is that
 MC-vertex operators $J_{\eta^{(k)}}$ (and resp. $\overline{J}_{\eta^{(k)}}$)
do not form a closed algebra by themselves for general values of parameters. If
this condition is not satisfied (which is generally the case), renormalization
requires that counterterms have to be added in such a way that this algebra closes. Consequently,
 the non-local charges are obviously no longer symmetries of this modified action.
A well-known
example is provided by simply-laced ATFTs with parameter $\beta$ :
for smaller values than the critical value $\beta^2=1$ , only tadpole 
renormalization is necessary. However, in the Kosterliz-Thouless region, ATFTs
are $\it not$ renormalizable \cite{Gri1} and exponential of $\it all$  the
 roots\,\footnote{For non-simply laced case, exponential operators associated 
 to
short or long roots have drastically different dimensions : one must
introduce fermions in such a way as to increase the conformal dimensions
of the exponential operators associated to short roots \cite{Gri2}.} must be added
in order to render them renormalizable. This situation will be relevant in
 further analysis. 

\section{Quantum algebraic structure and lowest-rank affine Lie algebras parametrization}

Known integrable models generally exhibit connections with Hopf algebras 
like Lie algebras and their quantum deformations. The aim of this section is to clarify in which sense the previous parameter space
of the multisine-Gordon model is restricted by imposing this kind of 
structure to the non-local conserved charge algebra. The principal motivation 
to find such a structure in the MSG models comes 
from the powerful framework that these algebras provide :  they can be sufficiently 
restrictive to allow a non-perturbative solution of the theory,  determining the $S$ matrices 
for instance \cite{8}.  From eqs. (\ref{A}), i.e.
(\ref{cont3}), (\ref{cont4}) and (\ref{contr5}) with solutions (\ref{change}), (\ref{change2}), it is convenient
to introduce the elements :
\beqa
&&C_{a,a}(\eta^{\vee},\eta) = 2, \ \ \ \ \ \ \ \  \ \ C_{a,a+\frac{n}{2}}(\eta^{\vee},\eta) = C_{a+\frac{n}{2},a}  = -2 \frac{A_a+B_a}{A_a-B_a} = -n_1^a  \nonumber\\
&&\label{system} C_{a,\frac{n}{2}-1}(\eta^{\vee},\eta) = C_{a+\frac{n}{2},n-1}  = C_{a+\frac{n}{2},\frac{n}{2}-1} = C_{a,n-1} = 2 \frac{B_a \delta_a}{A_a-B_a} = -n_2^a \\
&& C_{\frac{n}{2}-1,a}(\eta^{\vee},\eta) = C_{n-1,a+\frac{n}{2}}  = C_{\frac{n}{2}-1,a+\frac{n}{2}} = C_{n-1,a} = 2 \frac{B_a \delta_a}{A_{\frac{n}{2}-1}-B} = -n_3^a  \nonumber\\
&& C_{n/2-1,n-1}(\eta^{\vee},\eta) = C_{n-1,\frac{n}{2}-1}  = -2 \frac{A_{\frac{n}{2}-1}+B}{A_{\frac{n}{2}-1}-B} = -n_4 \nonumber
\eeqa
with $A_a = \alpha_a^2, B_a = \beta_a^2$, $\delta_a = \frac{m_a}{m_{\frac{n}{2}-1}}$, and :
\beqa
\label{defBeven} B = \sum_{a=0}^{\frac{n}{2}-2} B_a \delta_a^2.
\eeqa
As detailed in \cite{Pascal}, the braiding relations between 
$(J^{(k)}, {\overline J}^{(l)})$, $(J^{(k)}, {\overline H}^{(l)})$ and
 $({H}^{(k)}, {\overline J}^{(l)})$ arising from the non-local property of
 the currents are identical iff :
\beqa
\{n_1^a, n_4\} \in {\mathbb N} \ \ \ \mbox{and}\ \ \ \{n_2^a, n_3^a\} \in {\mathbb N}^* \ \ \ \mbox{for}\ \ \ a \in \{ 0,1,\dots,\frac{n}{2}-2\}\label{n}.
\eeqa
Under this assumption, to first order in CPT non-local conserved charges 
(\ref{mucur}) obey a $q$-deformed structure in MC-space \cite{Pascal}. Furthermore it is possible, in the usual complex space and independently of any representation, to define one other basis of non-local conserved currents; the expansion of the currents $J^{(k)}$ in MC-basis $\{ P_a, -P^2_a\}$ reads :
\beqa
J_{{\eta^{(k)}}^\vee} = \sum_{a=0}^{\frac{n}{2}-1} \Big[\sin(\alpha_a^{\vee} \phi_a) e^{\beta_a^{\vee} \varphi_a} P_{a+k} + \cos(\alpha_a^{\vee} \phi_a) e^{\beta_a^{\vee} \varphi_a} \big(-P^2_{a+k}\big)\Big]
\eeqa
with $\beta_{\frac{n}{2}-1}^{\vee} \varphi_{\frac{n}{2}-1} = -\sum_{a=0}^{\frac{n}{2}-2}\frac{m_a^{\vee}}{m_{\frac{n}{2}-1}^{\vee}} \beta_a^{\vee} \varphi_a$. Since $J_{\eta^{(k)\vee}}$ is conserved, then each one of its components (and any linear combinations of them) is a non-local conserved current, particulary :
\beqa
{\cal J}^{(a)} = e^{-i \alpha_a^{\vee} \phi_a + \beta_a^{\vee} \varphi_a},\\
{\cal J}^{(a+\frac{n}{2})} = e^{i \alpha_a^{\vee} \phi_a + \beta_a^{\vee} \varphi_a}\nonumber
\eeqa
and similarly for the antiholomorphic part. Non-local conserved charges
$({\cal Q}_a,{\overline{\cal Q}}_a)$ associated to these conserved currents
can then be obtained as in (\ref{mucur}). Analogously to the multicomplex case,
 these charges obey to 
a $q$-deformed algebra iff eqs. (\ref{system}), (\ref{defBeven}) with (\ref{n}) are satisfied. For 
$\alpha_a\in{\mathbb R}$ and $\beta_a\in i{\mathbb R}$, the resulting 
structure\,\footnote{For this parameter space, the normalization of the field 
is chosen such that ${\cal T}^{(a)}$ takes integer values \cite{Pascal}.} is nothing else than a ``parametrized'' quantum universal envelopping algebra 
${\cal U}_q({\hat{\cal G}})$ :
\beqa
{\cal Q}^{(a)}{\overline {\cal Q}}^{(b)} - q^{-C_{a,b}(\eta,\eta^{\vee})}_{a}{\overline {\cal Q}}^{(b)} {\cal Q}^{(a)} &=& \delta_{a,b}\frac{\lambda}{in\pi} \Big[ 1-q_{a}^{2{\cal{T}}^{(a)}}\Big].\label{Qalg2} 
\eeqa
where  $q_a=\exp(-i\frac{\pi}{2} C_{a,a}(\eta^{\vee},\eta^\vee))$ is the deformation
and ${\cal{T}}^{(a)}$ is a parametrized topological charge \cite{Pascal}. Here, coefficients $C_{a,b}(\eta,\eta^{\vee})=C_{b,a}(\eta^{\vee},\eta)$ given 
by eqs. (\ref{system}) correspond to the extended Cartan matrix elements of 
this ``parametrized'' $q$-deformed algebra. As we will see later, the type of
affine Lie algebra is encoded in the ratios of the parameters and the MC-algebra
structure (e.g. $A_a$, $B_a$ and $\delta_a$). As was shown in \cite{Pascal}, the matrix elements of the extended 
Cartan matrix for $n$ odd can be obtained similarly from eqs. (\ref{system}) by the 
substitution :
\beqa
n \to n+1 ;\ A_{\frac{n-1}{2}} \to 0; \ \delta_a \to 2 \delta_a,\label{substit}
\eeqa
while the last equation of (\ref{system}) does not appear. It is more convenient to formulate this substitution in terms of the $n_i^a$ so that solutions for the $n$ odd case can be directly read off from solutions of the even case (with $A_{\frac{n}{2}-1} = 0$). It reads :
\beqa
n \to n+1 ;\ n_4 \to -2;\ n_2^a \to \frac{n_2^a}{2};\ n_3^a \to \frac{n_3^a}{2},\label{sub1}
\eeqa
and (\ref{defBeven}) becomes for the odd case :
\bea
\label{defBodd} B = 4 \sum_{a=0}^{\frac{n-1}{2}} B_a \delta_a^2.
\eea
In the following, the system (\ref{system}) (and correspondingly for $n$ odd) is
solved\,\footnote{For completeness, we also report the cases 
$C_2^{(1)}$, $D_3^{(2)}$, $A_2^{(1)}$ and $A_3^{(1)}$ already obtained
 in \cite{Pascal}.}. First, using (\ref{system}), $A_a$,
$\delta_a$, $B_a$, $A_{\frac{n}{2}-1}$ are expressed in terms of $n_i^a$ and
$B$. Then, equation (\ref{defBeven}), (or (\ref{defBodd}) for $n$ odd) appear
as a further constraint that fix the $n_i^a$. Two
generic cases are now studied : $i$) no restriction
 $(\{\alpha_a\},\{\beta_a\})\neq 0$ , $ii$) restrictions on
parameters space.

\newpage

\centerline{\bf Case $n$ even without restriction}
\vspace{3mm}
The case with no restrictions corresponds to keeping $A_a \neq  0$, $B_a\neq 0$, $A_{\frac{n}{2}-1} \neq  0$. The general solution for the system (\ref{system}) reads (with $B < 0$) :
\beqa
\frac{A_a}{B_a} = -\frac{2+n_1^a}{2-n_1^a};&&\ \ A_{\frac{n}{2}-1} = -\frac{2+n_4}{2-n_4} B; \label{solevennorest} \\
B_a = \frac{n_3^a (2-n_1^a)}{n_2^a (2-n_4)} B;&&\ \ \delta _a = \frac{2 n_2^a}{2-n_1^a},\nonumber
\eeqa
with $n_1^a = 0$ or $1$, $n_2^a > 0$, $n_3^a > 0$ and $n_4 = 0$ or $1$.
Since $n_i^a \in {\mathbb N}^*$, $\delta_a$ are rational and the $m_a$
can then be integers. Here, $B$ is a negative real number, and due to 
(\ref{defBeven}), the $n_i^a$ must satisfy :
\beqa
\sum_{a=0}^{\frac{n}{2}-2} \frac{n_3^a n_2^a}{2-n_1^a} = \frac{2-n_4}{4}.     \label{cont1}
\eeqa
Since $\frac{n_3^a n_2^a}{2-n_1^a} \in \frac{1}{2} {\mathbb N}^{*}$, $n_4 = 0$ is
the only consistent case. It corresponds to the line $d/$ of Table 1.1.

\vspace{7mm}
\centerline{\bf Case $n$ odd without restriction}
\vspace{3mm}

The solution in this case can be obtained from (\ref{solevennorest}) with substitutions (\ref{sub1}) ($B<0$):
\bea
\label{soloddnorest}
\frac{A_a}{B_a} = -\frac{2+n_1^a}{2-n_1^a};\ \ B_a = \frac{n_3^a (2-n_1^a)}{4 n_2^a} B;\ \ \delta _a = \frac{n_2^a}{2-n_1^a},
\eea
with $n_1^a = 0$ or $1$, $n_2^a > 0$, $n_3^a > 0$, and the condition (\ref{defBodd}) :
\bea
\sum_{a=0}^{\frac{n-1}{2}-1} \frac{n_3^a n_2^a}{2-n_1^a} = 1 .   \label{cont1}
\eea
Bounds on the $n_i^a$ imply that  (\ref{cont1}) has solutions only if $n \le 5$. There are three solutions for $n=3$ corresponding to lines a/, b/, c/ of Table 1.1, and one for $n=5$ (line e/ of Table 1.1).

\vspace{0.3cm}
{\centering \begin{tabular}{|c|c|c||c|c|c||c|c|c|c||c|}
\hline
& $n$ & a & $n_{1}^{a}$ & $n_{2}^{a}$ & $n_{3}^{a}$ & $A_{a}/B_{0}$ & $A_{\frac{n}{2}-1}/B_{0}$ & $B_{a}/B_{0}$ & $\delta _{a}$ & Algebra\\
\hline
\hline
a/ & 3 & 0 & 0 & 1 & 2 & -1 & * & 1 & 1/2 & $C_2^{(1)}$\\
\hline
b/ & 3 & 0 & 0 & 2 & 1 & -1 & * & 1 & 1 & $D_3^{(2)}$\\
\hline
c/ & 3 & 0 & 1 & 1 & 1 & -3 & * & 1 & 1 & $A_2^{(1)}$\\
\hline
d/ & 4 & 0 & 0 & 1 & 1 & -1 & -1 & 1 & 1 & $(n_4=0)\ A_3^{(1)}$ \\
\hline
 &  & 0 & 0 & 1 & 1 & -1 &  & 1 & 1/2 & \vspace{-0.25cm}\\
e/ & 5 &  &  &  &  &  & * &  &  & $D_4^{(1)}$ \vspace{-0.25cm}\\
 &  & 1 & 0 & 1 & 1 & -1 &  & 1 & 1/2 &  \\
\hline
\end{tabular}\par}
\begin{center}
{\small \underline{Table 1.1} - Solutions without restriction, and the
 corresponding \\
hidden symmetry algebras.}
\end{center}

\vspace{5mm}

\centerline{\epsfig{file=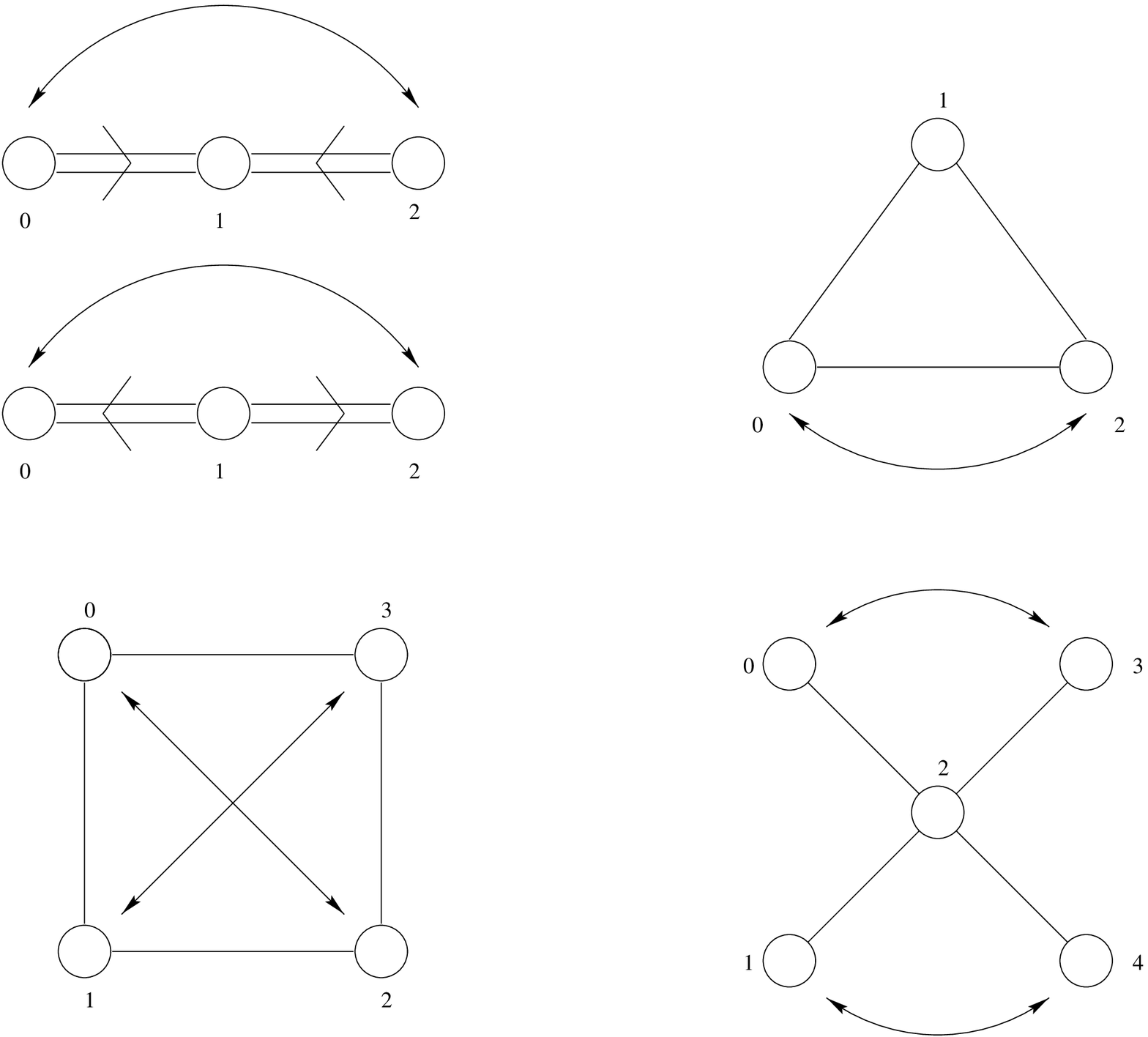,height=85mm,width=95mm}}
\vspace{5mm}

\centerline{\small \underline{Figure 1.1} - Small arrows symbolise the exchange under multicomplex conjugation}
\centerline{\small $a\rightarrow a+n/2$ (or $a\rightarrow a+(n+1)/2$ for $n$ odd) . The number of links is }
\centerline{\small $|C_{a,b}(\eta,\eta^\vee)||C_{a,b}(\eta^\vee,\eta)|$ where big arrows goes from $a$ to $b$ if $|C_{a,b}(\eta,\eta^\vee)|>|C_{a,b}(\eta^\vee,\eta)|$.}

\vspace{7mm}
\centerline{\bf Case $n$ even with restrictions}
\vspace{3mm}
We consider $s+1$ restrictions :
\bea
A_a &=& 0 \mbox{ for } a\in \{ 0,1,\dots ,s\}, \nonumber \\
A_a &\neq& 0 \mbox{ for } a \in \{s+1,\dots ,\frac{n-1}{2}-1\}\ ( \emptyset \mbox{ if } s = \frac{n-1}{2}-1). \label{def-rest-even}
\eea
The solution of (\ref{system}) for $a \le s$ can be obtained from (\ref{solevennorest}) by setting $n_1^a = -2$ :
\bea
\label{solevenwithrest}
A_a = 0;\ \ A_{\frac{n}{2}-1} = -\frac{2+n_4}{2-n_4} B;\ \  B_a = \frac{4 n_3^a}{n_2^a(2-n_4)} B;\ \ \delta_a = \frac{n_2^a}{2},
\eea
while (\ref{solevennorest}) still holds when $s+1 \le a \le \frac{n}{2}-1$. One can also set the restriction $A_{\frac{n}{2}-1} = 0$, in which case solutions of (\ref{system}) are obtained from (\ref{solevennorest}) and (\ref{solevenwithrest}) by setting $n_4 = -2$, (while $n_4 = 0$ or $1$ if $A_{\frac{n}{2}-1} \neq 0$). In all cases, one still have $n_2^a > 0, n_3^a > 0$. (\ref{defBeven}) yields the condition :
\bea
\label{C2}
\sum_{a=0}^{s} \frac{n_2^a n_3^a}{4} + \sum_{a = s+1}^{\frac{n}{2}-2} \frac{n_2^a n_3^a}{2-n_1^a} = \frac{2-n_4}{4}.
\eea
$\bullet$ For $A_{\frac{n}{2}-1} \neq 0$, this last equation admits solutions only if $n \le 6$ : For $n=4$, there are three solutions written on lines a/, b/, c/ of Table 1.2, and for $n=6$ there is one solution line n/ of Table 1.2, reported in Appendix A.\\
$\bullet$ For $A_{\frac{n}{2}-1} = 0$, the upper bound on $n$ is $10$. Solutions are written on lines d/ to m/ and o/ to r/ of Table 1.2 in Appendix A.

\vspace{7mm}
\centerline{\bf Case $n$ odd with restrictions}
\vspace{3mm}
In this case, one can only consider $s+1$ restrictions :
\bea
A_a &=& 0 \mbox{ for } a\in \{ 0,1,\dots ,s\}, \nonumber \\
A_a &\neq& 0 \mbox{ for } a \in \{s+1,\dots ,\frac{n-1}{2}\}\ ( \emptyset \mbox{ if } s = \frac{n-1}{2}). \label{def-rest-odd}
\eea
Solutions for $a \le s$ can be obtained from (\ref{soloddnorest}) by setting $n_1^a = -2$ :
\bean
A_a = 0;\ \ B_a = \frac{n_3^a}{n_2^a} B;\ \ \delta_a = \frac{n_2^a}{4},
\eean
while for $s+1 \le a \le \frac{n-1}{2}$ solution is given by 
(\ref{soloddnorest}), with $n_1^a = 0$ or $1$. In any case $n_2^a >0$ and $n_3^a >0$. The condition (\ref{defBodd}) becomes :
\bea
\sum_{a=0}^{s} \frac{n_2^a n_3^a}{4} +\sum_{s+1}^{\frac{n-1}{2}-1} \frac{n_2^a n_3^a}{2-n_1^a} = 1. \label{conditionmix}
\eea
Owing to the conditions on $n_i^a$, and the upper bound of $s$, (\ref{conditionmix}) have only solutions for $n \le 9$. Solutions are summarized on lines a/ to n/ of Table 1.3 in Appendix A.\\

To resume, imposing a quantum algebraic structure to the non-local charges  restricts the ($n-1$)-parameter space to a discrete set of one dimensional
 submanifolds. Each one is characterized by ratios of the parameters 
 ($\{\alpha_a\},\{\beta_a\}$) and ratios of $m_a$ which determinate 
 \ $n_1^a, n_2^a, n_3^a, n_4$ \  in eqs. (\ref{system}). Consequently, since
 (\ref{system}) are identified to Cartan matrix elements, any of these 
 submanifolds is in one-to-one correspondance with an affine Lie algebra
 (see tables. 1.1-1.3). We see that any lowest-rank affine
 Lie algebra and its dual appear for each fixed value of the multicomplex
  space of dimension $n$. This is easily understood from the substitutions
  (\ref{change}) and (\ref{change2}) in (\ref{Qalg2}), resulting from
 the duality property.
\section{Integrability in parameter space : affine Toda field theories and perturbed WZNW models}
Toda and affine Toda field theories are generally understood as the simple
Lie group extension of the Liouville and sine-Gordon models respectively. From
our previous results, we 
intend to show in this section that some of these theories, more precisely 
those based on lowest rank simply-laced algebras, can be generated starting
from a multicomplex number $e$ satisfying $e^{10}=-1$. Let us now consider the set of MC-vertex operators :
\beqa
y^{(l)}=n^{(l)}x^{(l)} \ \ \ \mbox{with}\ \ \ n^{(l)} = \sum_{a=0}^{\frac{n}{2}-1} n_a (-P^2_{a+l}). 
\eeqa
The pseudo-norm associated to these operators is :
\beqa
\label{toda-unimod}
||y||^{m} = || n x ||^m = \prod_{a=0}^{\frac{n}{2}-1} n_a^{2m_a}\label{pnorm}
\eeqa
as $||x ||^{m} = 1$. It is then straightforward to see that among the models which can be built in terms
 of these MC-vertex operators, the ones with action (for $n$ even) :
\beqa
{{\cal{A}}}^{(n|m)}(\eta)= \frac{1}{4\pi}\int d^2z \partial_z\Phi\partial_{\overline z}\Phi + \frac{\lambda}{n\pi}\int d^2z \sum_{k=0}^{n-1} n^{(k)} x^{(k)} \label{action7}  
\eeqa
possess exactly the same underlying algebraic structure than those considered
in previous sections. Except for the explicit expression of the non-local
 conserved charges and r.h.s of (\ref{Qalg2}), which 
are modified by the presence of the extra-parameters $n_a$ appearing 
through the changes in the non-diagonal terms :
\beqa
H^{(k)} \rightarrow n^{(k)}H^{(k)},
\eeqa
in (\ref{mucur}), the whole analysis concerning the non-local conserved charge algebraic 
structure is preserved. In the usual complex space 
(\ref{action7}) for $n$ even writes :
\beqa
{{\cal{A}}}^{(n|m)}(\eta)= \frac{1}{4\pi}\int d^2z \partial_z\Phi\partial_{\overline z}\Phi + \frac{2\lambda}{n\pi}\int d^2z \sum_{a=0}^{\frac{n}{2}-1} n_a \cos(\alpha_a \phi_a) e^{\beta_a \varphi_a} \label{action4}  
\eeqa
with $ \beta_{\frac{n}{2}-1} \varphi_{\frac{n}{2}-1} = -\sum_{a=0}^{\frac{n}{2}-2}\frac{m_a}{m_{\frac{n}{2}-1}} \beta_a \varphi_a$. This action can be generally put into the form :
\beqa
{{\cal{A}}}^{(n|m)}(\eta)=\frac{1}{4\pi}\int d^2z \partial_z\Phi\partial_{\overline z}\Phi + \frac{\lambda}{n\pi}\int d^2z \Big[\sum_{a=0}^{n-1}n_{a}e^{\beta_0{\bf r}_{a}.\Phi}\Big]\label{action3}.
\eeqa
with \ $n_a=n_{a+\frac{n}{2}}$\  for \ $a\in\{0,...,\frac{n}{2}-1\}$, and 
where we introduce some ``parametrized'' roots ${\bf r}_a$, reported in Appendix B. 

Similarly, to obtain the multisine-Gordon action for $n$ odd \cite{multisine,Pascal},
 we do successively the substitutions (\ref{substit}) in (\ref{action7}), (\ref{action4}) and
(\ref{action3}), then change
 $n_{\frac{n-1}{2}}\rightarrow n_{\frac{n-1}{2}}/2$.  For $n$ odd,
 ``parametrized'' roots are deduced using the same method. As we see, the
form of the action recalls the standard one of affine Toda field 
theories.
Taking the Lagrangian based on the multisine potential, we expand the 
interaction term around the minimum at $\Phi={\bf 0}$ :
\beqa
\sum_{a=0}^{n-1}n_{a}e^{\beta_0{\bf r}_{a}.\Phi} \sim \sum_{a=0}^{n-1}n_{a} + \beta_0\sum_{a=0}^{n-1}n_{a}{\bf r}_{a}^i\Phi^i &+& \frac{\beta_0^2}{2!}\sum_{a=0}^{n-1}n_{a}{\bf r}_{a}^i{\bf r}_{a}^j\Phi^i\Phi^j \\
&+& \frac{\beta_0^3}{3!}\sum_{a=0}^{n-1}n_{a}{\bf r}_{a}^i{\bf r}_{a}^j{\bf r}_{a}^k\Phi^i\Phi^j\Phi^k +...\nonumber
\eeqa
Stabilization of the classical vacuum implies cancellation of the linear term. 
Using (\ref{action4}), for each $n$ even or $n$ odd (with (\ref{substit})) case this is ensured iff :
\beqa
\frac{n_a}{n_{E(\frac{n+1}{2})-1}} = \frac{m_a}{m_{E(\frac{n+1}{2})-1}}\ \ \ \mbox{for} \ \ a\in\{0,...,E(\frac{n+1}{2})-2\},\label{kac}
\eeqa
where $E(x)$ stands for the integer part of $x$. Consequently, we obtain the linear relation among
 the ``parametrized'' roots : 
\beqa
\sum_{a=0}^{n-1}m_{a}{\bf r}_{a}=0.\label{rel}
\eeqa
Such kind of relation is characteristic of ATFTs. One of the roots is generally
identified with the negative of the highest root of the finite Lie algebra 
${\cal{G}}$ (with rank $n-1$) considered and the set of different numbers
 $m_a$ in (\ref{rel}) are proportional to the Kac labels. From eq.
 (\ref{kac}), it is interesting to see that ratios of these labels are
 completly defined by the kind of MC-algebra choosed.
 Furthermore, using eq. (\ref{kac}), the dual Kac labels associated with the
 dual ATFTs transform as in eq. (\ref{change2}). Under these considerations,
to each ATFT (and its dual one) corresponds the pseudo-norm (\ref{pnorm}). As the ratios of ${m_a}$ take rational values (see the
previous section), there exists one faithful representation
 $\pi$ \cite{Pascal}, given by $(n\times n)$ dimensional diagonal matrices:
\beqa
\pi\big[P_{a}\big]= Diag(0,...,0,i,0,...,0,-i,0,...,0)\ \  \  \ \mbox{for} \ \ a\in\{0,...,n/2-1\},\label{rep}
\eeqa
where \  $i$ \ (resp. $-i$) \ is in the $a$\ (resp.\  $n-1-a$)\ position, e.g. :
\beqa
&&{(\pi\big[P_{a}\big])}_{jj}= (\pi\big[{P_{a}}^{\dagger}\big])_{n-1-j,n-1-j}\ \ \mbox{where $\dagger$ denotes the ordinary hermitian}\nonumber\\
    && \ \mbox{conjugate, for}\ \ \  (a,k)\in\{0,...,n/2-1\}\ \mbox{and}\ j\in\{0,...,n-1\}.\nonumber
\eeqa
For instance, consider the multisine-Gordon model in multicomplex dimension
 $n=3$ denoted by \ $MSG_{(3|m)} (\alpha_0,\beta_0)$, \ and its dual denoted by
 \ $MSG_{(3|m^\vee)} (\alpha_0^\vee,\beta_0^\vee)$. From eqs. (\ref{change2})
  with substitutions (\ref{substit}), we have :
\beqa
\frac{m_0^\vee}{m_1^\vee} = \frac{m_1}{m_0}\left[ \frac{1}{2-n_1^0} \right].
\eeqa 
As \ $m = 2m_0 + m_1$ \ and similarly \ $m^\vee = 2m_0^\vee + m_1^\vee = m +\Delta$\ \ with \ $\Delta = m_1-m_0n_1^0$, we obtain three solutions :\\

\ \ \ \ $\bullet$  \  $n_1^0 = 0$ : for \ $m_0=m_1=1 \ (\Delta =1)\ \rightarrow \ m=n=3$ we obtain the model denoted ${\cal{A}}^{(3|3)}(\beta_0,i \beta_0)$ and for \ $2m_0^\vee=m_1^\vee=2 \ \rightarrow \ n=3\neq m^\vee=4$\  we obtain the model denoted \ ${\cal{A}}^{(3|4)}(- \frac{1}{\beta_0},-\frac{i}{ \beta_0})$. These two models describe respectively the non-simply laced $D_3^{(2)}$ and $C_2^{(1)}$ ATFTs.\\

\ \ \ \ $\bullet$ \ $n_1^0 = 1$ : for \ $m_0=m_1=1 \ (\Delta =0)\ \rightarrow \ m=m^\vee=n=3$ we obtain the model denoted ${\cal{A}}^{(3|3)}(\sqrt{\frac{3}{2}} \beta_0,\frac{i}{\sqrt{2}} \beta_0)$. It is self-dual and describes the simply laced $A_2^{(1)}$ ATFT.\\
These three models are all generated by a fundamental multicomplex number $e$ satisfying $e^3=-1$ (or its dual $e^\vee$) which possesses the ($m\times m$) matrix representation $\pi'$ : 
\beqa
&{\cal{A}}^{(3|m)}(\alpha_0,\beta_0) \mbox{\ with respect to\  } \pi'[e] = &\hspace{-0.3cm} \mbox{Diag} [\ \underbrace{e^{i\frac{\pi}{3}} \dots e^{i\frac{\pi}{3}}  },\ \  \underbrace{ -1 \dots -1}  ,\ \ \underbrace{ e^{-i\frac{\pi}{3}} \dots e^{-i\frac{\pi}{3}} } \ ]\nonumber \\
&& \mbox{\ \ \ \ \ \ ($m_0$ times) \  ($m_1$ times)\ \ \  ($m_0$ times)}\nonumber \\
&\hspace{-1cm} \updownarrow \mbox{ dual with } & \nonumber\\ \nonumber\\
&{\cal{A}}^{(3|m^\vee)}(\alpha_0^\vee,\beta_0^\vee) \mbox{\ with respect to\ } \pi'[e^\vee] = &\hspace{-0.3cm} \mbox{Diag} [\ \underbrace{e^{i\frac{\pi}{3}} \dots e^{i\frac{\pi}{3}}  },\ \  \underbrace{ -1 \dots -1}  ,\ \ \underbrace{ e^{-i\frac{\pi}{3}} \dots e^{-i\frac{\pi}{3}} } \ ] .\nonumber \\
&& \mbox{\ \ \ \ \ ($m_0^{\vee}$ times)\ \  ($m_1^{\vee}$ times)\ \ \  ($m_0^{\vee}$ times)}\nonumber 
\eeqa

\vspace{2mm}

Consider now the multisine-Gordon model for $n=4$. The $q$-deformed structure
 of the non-local charge algebra is ensured for the choices 
 \ $n_1^0 = 0$ and $n_4=0$. The resulting model ${\cal{A}}^{(4|4)}$
 corresponds to the simply laced $A_3^{(1)}$ ATFT. Since its hidden symmetry 
 is self-dual under transformations (\ref{change}) and (\ref{change2}), the
 weak-strong coupling regimes (with respect to the parameter $\beta_0$) of the two dual 
 actions are identical \cite{Ari}. However, while the fundamental multicomplex number
 representation associated with one model is :
\beqa
&&\hspace{-1.3cm}{\cal{A}}^{(4|m)}(\alpha_0,\alpha_1;\beta_0) \mbox{\  generated by :\ } \nonumber \\
&&\pi'[e] =  \mbox{Diag} [ \underbrace{e^{i\frac{\pi}{4}} \dots e^{i\frac{\pi}{4}}},\ \underbrace{e^{i\frac{3\pi}{4}} \dots e^{i\frac{3\pi}{4}}}, \  \underbrace{e^{-i\frac{3\pi}{4}} \dots e^{-i\frac{3\pi}{4}}} ,\ \underbrace{ e^{-i\frac{\pi}{4}} \dots e^{-i\frac{\pi}{4}}} ] ,\nonumber \\
&& \mbox{ \ \ \ \ \ \ \ \ \ \ \ \ \ \ ($m_0$ times),  ($m_1$ times), \ \  ($m_1$ times), \ \ \ \ ($m_0$ times) }  \nonumber 
\eeqa
its dual representation, associated with the dual model is obtained from 
\ $\frac{m_0^\vee}{m_1^\vee} = \frac{m_1}{m_0}$, which corresponds to :
\bean
&& \hspace{-1cm} {\cal{A}}^{(4|m^\vee)} (\alpha_0^\vee,\alpha_1^\vee;\beta_0^\vee) \mbox{ \ generated by :\ } \nonumber \\
&& \pi'[e^\vee] =  \mbox{Diag} [  \underbrace{e^{i\frac{\pi}{4}} \dots e^{i\frac{\pi}{4}} }\ \ ,\ \ \underbrace{e^{i\frac{3\pi}{4}} \dots e^{i\frac{3\pi}{4}}}\ \ ,\  \underbrace{ e^{-i\frac{3\pi}{4}} \dots e^{-i\frac{3\pi}{4}} }\ ,\  \underbrace{e^{-i\frac{\pi}{4}} \dots e^{-i\frac{\pi}{4}} } \  ] .\nonumber \\
&& \mbox{\ \ \ \ \ \ \ \ \ \ \ \ \ \ \  \ecran{m_0^\vee=m_1}{times},\ecran{m_1^\vee=m_0}{times}\ ,\ \ecran{m_1^\vee=m_0}{times},\ \ecran{m_0^\vee=m_1}{times} } \nonumber
\eean
Then, self-duality of $A_3^{(1)}$ translates into the following 
exchange : 
\beqa
e \longleftrightarrow e^\vee = -\frac{1}{e}
\eeqa
in the multicomplex space.

Action (\ref{action7}) in the
multicomplex space (and its dual multicomplex) and action (\ref{action3}) 
in the usual complex space provide a unified representation of all  
lowest-rank affine Toda field theories. If the perturbation is relevant, 
no new operator will be generated under renormalization flows. Non-local 
MC-currents are conserved to all order in CPT and consequently quantum duality 
is satisfied. Tables 1.1-1.3 give the list of 
ATFTs described in this formalism whereas the exchange of the node under multicomplex conjugation in the corresponding Dynkin diagrams is depicted in fig. 1.1-1.3.

However, if the perturbation becomes marginal, MC-vertex operators associated
to non-simple roots are generated under renormalization. For instance, let us
 consider the simply-laced cases ($A_2^{(1)}$, $A_3^{(1)}$ and $D_4^{(1)})$ 
in Table 1.1. The perturbation is marginal if $\beta_0^2 = 1$. 
 The perturbing (self-dual) operator in MC-space can be
written in the standard complex space as a current-current perturbation of a 
level-one WZNW model. We obtain respectively $su(3)$, $su(4)$
 and $so(8)$ current-current perturbations\,\footnote{The 
 resulting action is of the Gross-Neveu type, obtained from the bosonization
 of the ${\cal G}$-invariant Gross-Neveu models.}\cite{8}. This suggests
 a possible representation of WZNW or Gross-Neveu models in MC-space.

\section{MC-algebra and perturbed conformal field theories}
Instead of considering action (\ref{action}) (or its dual counterpart to first order in 
CPT) as a perturbed $(n-1)$ free field CFT, we can proceed differently.
Let us consider the MSG potential \ \ $\sum_{l} x^{(l)}$ \ \ for \ \  
$l\in\{0,...,n-1\}$. If we truncate this potential by supressing one of the
 MC-operators, say \ \ $x^{(0)}$, 
the resulting form is no longer invariant under multicomplex 
conjugation. However, conformal invariance can be realized by adding some specific MC-charge at infinity coupled to the fundamental field of the theory. $x^{(0)}$ is
 then identified as the perturbation in MC-space. To show that, we define the holomorphic part of the MC-stress-energy tensor : 
\begin{eqnarray}
T(z)=-\frac{1}{2}(\partial\Phi)^2 + {\sqrt 2}\beta_0 {\bf Q}.\partial^2\Phi , \label{stress}
\end{eqnarray}
where ${\bf Q}\in {\mathbb{MC}}_{(n|m)}$ and similarly for the antiholomorphic part. For further convenience\,\footnote{It is also possible to expand ${\bf Q}$ over generators $P_b$ instead of $-P_b^2$. In any case, the central charge takes real values.}, we write ${\bf Q}$ as :
\beqa
{\bf Q}= \sum_{b=0}^{n/2-1}{\bf Q}_{b}(-P_{b}^2).\label{Qb}
\eeqa  
For $n$ even, using the representation (\ref{rep}) we define the $n/2$ projections $\pi_a$ over the usual complex space with : 
\begin{eqnarray}
\pi_a(x^{(0)})=\big(\pi[x^{(0)}]\big)_{aa}=e^{\beta_0{\bf r}_a.\Phi}\ \ \ \ \mbox{for}\ \ a\in\{0,...,n/2-1\}.\label{r} 
\end{eqnarray}
From eq. (\ref{r}), we see that although the MSG Lagrangian representation does {\it not} depend on the choice of a particular projection, the Lagrangian representation in the usual complex space of the CFT part (i.e. equivalently the expression of the perturbation in the complex space) depends on this choice. In fact, due to the permutation symmetry of the ``parametrized'' roots ${\bf r}_a$ for $a\in\{0,...,n/2-2\}$ (see Appendix B)
the differences between all possible perturbations 
reduce to two distinct cases. In the first case, we use the projection\,\footnote{In fact for $n$ even (and similarly for $n$ odd) all projections $\pi_a$ for $a\in\{0,...,n/2-2\}$
are equivalent under multicomplex conjugation up to a permutation of the roots. For $n$ odd, using substitutions
(\ref{substit}), it is obvious to see that the nodes associated respectively to 
$n/2-1$ and $n-1$ ``collapse'' together as $\alpha_{n/2-1}\rightarrow 0$.} \ \ $\pi_{n/2-1}$ and the corresponding 
QFT is denoted \ $\cal{P}$. In the second case, we proceed similarly and use the projection \ \ $\pi_{0}$. 
The corresponding QFT is then denoted $\overline{\cal{P}}$. 
Each charge ${\bf Q}_{b}$ is then associated to the CFT obtained from the projection $\pi_b$. Consequently, an appropriate choice of the charge 
{\bf Q}, e.g. of the set $\{{\bf Q}_{b}\}$ ensures {\it simultaneously} the conformal invariance of {\it all} CFTs. From these previous remarks, it reduces to study only $\cal{P}$ and $\overline{\cal P}$, i.e. to calculate ${\bf Q}_{n/2-1}$ and ${\bf Q}_{0}$. It translates into a condition on the conformal dimensions $\Delta_{n/2-1}$ and $\Delta_{0}$ of the vertex operators :
\bea
\label{Delta=1}
\Delta_{b} \Big( e^{\beta_0 {\bf r}_a .\Phi} \Big)=1\ \ &&\mbox{for}\ \  {\cal P}\ (\pi_{n/2-1}),\ \  \mbox{i.e.} \ \ \mbox{for all} \ \ a \in \{ 0,...,\frac{n}{2}-2,\frac{n}{2}, \dots, n-2\}, \nonumber\\
\mbox{or} \ &&\mbox{for}\ \  {\overline{\cal P}}\ (\pi_{0}),\ \ \ \ \ \ \mbox{i.e.} \ \ \ \mbox{for all} \ \ a\in \{ 1,...,n-1\}.
\eea
Using eq. (\ref{stress}) and (\ref{Qb}), the holomorphic conformal dimension of each vertex 
operator is : 
\bea
\label{conf-weight}
\Delta_b \Big( e^{\beta_0  {\bf r}_a .\Phi} \Big)= - \frac{\beta_0^2}{2}  {\bf r}_a^2 + {\sqrt 2} \beta_0 {\bf Q}_b. {\bf r}_a \ .
\eea
For further convenience and by analogy with ATFTs approach, let us introduce :
\bea
\label{Q/beta0}
{\bf Q}_b = \frac{1}{\sqrt 2}\big[\beta_0{\bf \rho}_b + \beta_0^{\vee}{\bf \rho}_b^{\vee}\big]
\ \  \ \ \ \mbox{where} \ \ \ \
{\bf \rho}_b =  \sum_{\{c\}}{\bf \omega}_{c;b} \ \ \  \mbox{and} \ \ \  {\bf 
\rho}_b^{\vee} = \sum_{\{c\}}{\bf \omega}_{c;b}^{\vee}  
\eeqa
with \ $c\neq n/2-1$ for $\cal{P}$ and \ $c\neq 0$ for
 $\overline{\cal{P}}$. Eqs. (\ref{Delta=1}) can be satisfied
  if \ ${\bf \omega}_{c;b},{\bf \omega}_{c;b}^{\vee}$\ are choosed to obey :
\bea
{\bf \omega}_{c;b}^\vee.{\bf r}_a = \delta_{ac}\frac{1}{\beta_0\beta_0^\vee}.\label{system-co-root}
\eea
Similarly, the same approach can be applied to the $n$ odd case and the
 corresponding results are obtained using the substitutions (\ref{substit}).
 In each case ($\cal{P}$ or $\overline{\cal P}$) 
and any value of $n$, it is straightforward to compute the MC-central charge of the 
conformally invariant part:
\beqa
\label{centr-charge}
c=n-1+24|{\bf Q}|^2.
\eeqa 
We notice that expression (\ref{Q/beta0}) is self-dual under the 
duality transformation (\ref{change}). From the above analysis, we have computed
the ``parametrized'' central charges $c_b=\pi_b(c)$ for $\cal P$ and $\overline{\cal P}$, expressed in terms of $n$, $(m_a,m_a^{\vee})$, $(m,m^\vee)$,
 $\{{\bf r}_a\}$ and $\{\beta_a\}$. The ``parametrized'' co-weights are given in
 Appendix B and their associated weights are obtained using transformations
 (\ref{change}). Using eq. (\ref{centr-charge}) with eqs. (\ref{Qb}) and (\ref{Q/beta0}) we obtain :
\begin{itemize}
\item For\ \ ${\cal P}$ :
\beqa
&& c_{n/2-1}= n-1 + 12\big[\Gamma^{(1)}_{n;n/2-1} + \Gamma^{(2)}_{n;n/2-1}\big]\ \ \ \ \ \ \mbox{for $n$ even},\\
&& c_{(n-1)/2}= n-1 + 12\big[\Gamma^{(1)}_{n;(n-1)/2}\big]\ \ \ \ \ \ \ \ \ \ \ \ \ \ \ \ \ \ \mbox{for $n$ odd}.\nonumber
\eeqa
\item For\ \ ${\overline{\cal P}}$ and $n$ even :
\beqa
c_0= n-1 + 12\big[\Gamma^{(1)}_{n;0} + \Gamma^{(2)}_{n;0} +
\Gamma^{(3)}_{n;0}\big]
\eeqa
\end{itemize}
and where we define for $n$ even :
\beqa
&&\Gamma^{(1)}_{n;b}=\sum_{ \hspace{-0.1cm}   {~\shortstack{ $_{a=0}$\\ $_{a\ne b}$ } }    }^{n/2-2} \Big[\big( \frac{{\bf r}_a^2}{2} \beta_0 + \frac{1}{\beta_0} \big)^2 \big( \frac{\beta_0}{\beta_a}\big)^2\Big],\nonumber \\ 
&&\Gamma^{(2)}_{n;b}=-\big(\frac{\sum_{a=0}^{n/2-1} m_a {\bf r}_a^2}{2m_{b}} \beta_0 + \frac{m}{2m_{b}} \frac{1}{\beta_0} \big)^2 \big(\frac{\beta_0}{\alpha_{b}}\big)^2,\label{Gamma}\\ 
&&\Gamma^{(3)}_{n;0}=\big(\frac{\sum_{a=1}^{n/2-1} m_a {\bf r}_a^2}{2m_{0}} \beta_0 + \frac{m-2m_0}{2m_{0}} \frac{1}{\beta_0} \big)^2.\nonumber 
\eeqa
For \ \ ${\overline{\cal P}}$ and $n$ odd, we use substitution (\ref{substit}).

The conformal dimension of each vertex operator, exept the
perturbation, is one. Then, their integrals appear naturally as ``screening
charges'' whereas the conformal dimension of the perturbation for any 
projection $\pi_b$ is :  
\beqa
\Delta_b \left( \pi_b(x^{(0)}) \right) = \Delta_b \left( e^{\beta_0 {\bf r}_b . \Phi} \right) =
 1- \left[ \frac{m^{\vee}}{m_b^{\vee}}
 \left(\beta_0^2\frac{{\bf r}_b^2}{2} \right) + \frac{m}{m_b} \right]\label{delta}. 
\eeqa
For \ $(\{\alpha_a\}\in {\mathbb R}$, $\{\beta_a\}\in{i\mathbb R})$, 
the perturbation is relevant ($\Delta_b<1$) or marginal ($\Delta_b=1$) iff:
\beqa
(\beta_0^{\mathbb R})^2 \frac{{\bf r}_{b}^2}{2} \le \frac{\ \ \frac{m}{m_{b}}\ \ }{\ \ \frac{m^{\vee}}{m_{b}^{\vee}}\ \ }\label{relev}.
\eeqa
The renormalizability property of the model is then encoded in the MC-algebra.

As we saw previously, depending on the dimension, the 
specific structure of the MC-algebra (e.g. the values of $m_a$) 
and ratios of the parameters, the resulting model possesses a $q$-deformed 
symmetry. For each projection which leads to $\cal P$ or
 $\overline{\cal P}$, we have computed for the ratios reported in Table 1.1 the
central charge associated with the truncated MSG model, i.e. without 
the perturbing term $x^{(0)}$. For instance, for projection
 $\cal P$ for case a/ and b/ the truncated model is identified
to an $su(2)\otimes su(2)$ theory (two decoupled Liouville
 theories), whereas case e/ corresponds to an $so(4)\otimes so(4)$
theory (four decoupled Liouville theories). For projection $\overline{\cal P}$,
Toda field theories are obtained.  For any projection, case c/ and d/ are
respectively identified with $A_2$ and $A_3$ Toda field theories. Using the results 
reported in Table 1.1, it can be checked that agreement is obtained with the
results found in \cite{Hollo}. However, for imaginary values of the parameters 
$\beta_a$, a truncation of the Hilbert space is necessary in order to obtain a 
unitary theory. As ratios are fixed, the values of $\beta_0$ are fixed in each 
case, corresponding to a quantum group restriction of the model. This last restriction
corresponds to points in parameter space. 

For generic values of the parameters the situation is much more complicated.
Although the central charge associated with the truncated MSG model can be 
computed, it is not clear if unitary\,\footnote{However, many statistical 
systems do not satisfy the unitarity condition (nonself intersecting polymer
 chains, magnetics with stochastic interactions, etc...).} representations of the Virasoro algebra
always exist \cite{Fei}. However, for ${\bf r}_a^2=2$ and $a\in \{0, \dots, \frac{n}{2}-1\}$ (which is
the standard convention for simply laced affine Lie algebra), we notice
that any ``truncation'' (${\cal P}$ or ${\overline{\cal P}}$) of the basis
 of MC-vertex operators in action (\ref{action}) leads to a self-dual CFT (ratios of 
parameters are fixed) and condition (\ref{relev}) becomes ${(\beta_0^{\mathbb R})}^2\leq 1$. 

\section{Concluding remarks}
Consider the MSG model with action ${\cal{A}}^{(n|m)}(\eta)$ generated by the MC-algebra 
${\mathbb M \mathbb C}_{(n|m)}$. The parameter space can be described as follow. 
In the deep ultraviolet, it exists a discret set of one-dimensional
submanifolds \ ${\cal S}_0=(\{\alpha_a^{(0)}\},\{\beta_a^{(0)}\})$ (see Tables 1.1-1.3)
which corresponds to a scale invariant theory, i.e a CFT. This CFT with MC-central charge (\ref{centr-charge}) possess a
Lagrangian representation in MC-space in terms of a truncated basis of MC-vertex operators,
  $\sum_{l=1}^{n-1}x^{(l)}$ for instance.
In usual complex space, it admits two unequivalent Lagrangian representations $\cal P$ and
$\overline{\cal P}$. A parametrized central charge 
$c_b=c_b(\{\alpha_a^{(0)}\},\{\beta_a^{(0)}\})$ for $b=0$ or $E((n+1)/2)-1$ was obtained for each representations,
describing the fixed points of different known CFTs like decoupled Toda or
Toda field theories.

We are now interested in the neighbourhood region of these fixed points where we loose
the scale invariance of the model. As the CFT in MC-space is not invariant under 
multicomplex conjugation, a natural perturbation is then provided by imposing
the MC-conjugation symmetry to the resulting model. Indeed this 
perturbation, say $x^{(0)}$, does not correspond to an {\it arbitrary}
deformation of the CFT. The perturbed model always corresponds to a
one-parameter family of integrable massive QFTs. For 
${\cal S}^+_0=(\{\alpha_a^{(0)}\}\in{\mathbb R},\{\beta_a^{(0)}\}\in i{\mathbb R})$,
the model is identified to lowest-rank imaginary coupling ATFTs which 
possess soliton solutions. However, unitarity is only assumed
at specific points on ${\cal S}^+_0$. The theory gets truncated and the hidden
symmetry acts as a parametrized quantum group with the deformation parameters
 $q_b(\{\alpha_a^{(0)}\},\{\beta_a^{(0)}\})$ being a root of unity. For
 ${\cal S}^-_0=(\{\alpha_a^{(0)}\}\in i{\mathbb R},\{\beta_a^{(0)}\}\in {\mathbb R})$, 
lowest-rank real coupling ATFTs are obtained.

Exept in the Kosterlitz-Thouless region of ${\cal S}_0$, no new terms are
generated under the renormalization group flow. Non-local MC-conserved currents 
$\{J_{\eta^{(k)\vee}},{{\overline J}_{\eta^{(k)\vee}}}\}$ exist which are 
conserved to {\it all} orders in CPT. There, the MSG model (\ref{action2}) with 
(\ref{ver}) admits a dual Lagrangian representation with action 
${\cal{A}}^{(n|m^\vee)}(\eta^\vee)$, generated by the dual multicomplex algebra
${\mathbb M \mathbb C}_{(n|m^\vee)}$. The weak coupling regime of one MSG and 
the strong coupling regime of its
dual are simply related through the parameter exchange 
$\beta_0\leftrightarrow1/\beta_0$. 
  
The neighbourhood region of ${\cal S}_0$ $(\lambda<<1)$ is much more complicated but a few 
remarks can be done from our previous analysis. First, the non-local
currents (and dual ones) are still conserved, at least to first order
in CPT, for a larger parameter space ${\cal S}_1$ described by eqs. (\ref{cont3}), (\ref{cont4}), (\ref{contr5}).
It is easy to see that the symmetry group of ${\cal S}^+_1$ (and similarly for ${\cal S}^-_1$ with the opposite signature) is
 the pseudo-rotational group $SO(\frac{n}{2}-R,\frac{n}{2}-1)$ for $n$ even and $SO(\frac{n-1}{2}-R,\frac{n-1}{2})$ for $n$ odd.
However, the conserved charges (\ref{mucur}) do not satisfy the previous
$q$-deformed symmetry since braiding relations between MC-operators strongly 
depend on the ratios of parameters. A less restrictive algebraic structure, 
like more general Hopf algebras, may exist. It is known that integrable models
can be generated by some more general algebras than the quantum Lie algebras :
the integrability condition is dictated by the Yang-Baxter equation itself
\cite{Ku}. Secondly, the non-local conserved currents
does {\it not} necessary form a closed algebra. Then, renormalization group flow
may generated counterterms which can spoiled the current conservation to
higher order in CPT.

To conclude, we would like to mention that {\it all} the integrable (dual-)QFTs 
 in \cite{multisine,4} and here are generated by the MC-algebra
${\mathbb M \mathbb C}_{(n|m)}$ for specific values of $(n,m)$. The difference
 between them simply reduces to the parameter space. Particulary, the non-local 
 conserved charges (\ref{mucur}), if they satisfy a $q$-deformed 
algebraic structure \cite{Pascal}, can be simply expressed in terms of 
a Lie algebra based on the field ${\mathbb M \mathbb C}_{(n|m)}$
(multicomplex are of characteristic 0 as for the usual complex numbers), where
parameters $(\alpha_a,\beta_a)$ play the role of deformations.
Furthermore, it is believed that more general MC-algebras exist \cite{4}.
Many other QFTs should then be described within this formalism.

\paragraph*{Aknowledgements}
We are very grateful to C. Ahn, J.L. Kneur, A. Neveu, M. Rausch de Traubenberg, F. Smirnov, G. Tak\'acs
and J. Thierry-Mieg for useful discussions.  P.B. thanks for the hospitality of KIAS
where part of this work was done. Work supported in part by the EU under contract 
ERBFMRX CT960012.

\newpage
\centerline{\bf \large Appendix A}
\vspace{0.2cm}
{\centering \begin{tabular}{|c|c|c|c||c|c|c||c|c|c|c||c|}
\hline 
& $n$ & $R$ &  a & $n_{1}^{a}$ & $n_{2}^{a}$ & $n_{3}^{a}$ & $A_{a}/B_{0}$ & $A_{\frac{n}{2}-1}/B_{0}$ & $B_{a}/B_{0}$ & $\delta_{a}$ & Algebra\\
\hline 
\hline 
a/ & 4 & 1 & 0 & * & 2 & 1 & 0 & -1 & 1 & 1 &($n_4=0$) $C_2^{(1)}$ \\
\hline 
b/ & 4 & 1 & 0 & * & 1 & 2 & 0 & -1/4 & 1 & 1/2 &($n_4=0$) $D_3^{(2)}$ \\
\hline 
c/ & 4 & 1 & 0 & * & 1 & 1 & 0 & -3/4 & 1 & 1/2 &($n_4=1$) $A_2^{(1)}$ \\
\hline 
d/ & 4 & 2 & 0 & * & 4 & 1 & 0 & 0 & 1 & 2 & $A_2^{(2)}$ \\
\hline 
e/ & 4 & 2 & 0 & * & 1 & 4 & 0 & 0 & 1 & 1/2 & $A_2^{(2)}$ \\
\hline 
f/ & 4 & 2 & 0 & * & 2 & 2 & 0 & 0 & 1 & 1 & $A_1^{(1)}$ \\
\hline 
 &  &  & 0 & * & 1 & 2 & 0 &  & 1 & 1/2 & \vspace{-0.25cm}\\
g/ & 6 & 2 &  &  &  &  &  & 0 &  &  & $A_5^{(2)}$ \vspace{-0.25cm}\\
 &  &  & 1 & 0 & 1 & 1 & -1/4 &  & 1/4 & 1 & \\
\hline 
   &  &  & 0 & * & 2 & 1 & 0 &  & 1 & 1 & \vspace{-0.25cm}\\
h/ & 6 & 2 &  &  &  &  &  & 0 &  &  & $B_3^{(1)}$ \vspace{-0.25cm}\\
   &  &  & 1 & 0 & 1 & 1 & -1 &  & 1 & 1 & \\
\hline 
   &  &  & 0 & * & 1 & 1 & 0 &  & 1 & 1/2 & \vspace{-0.25cm}\\
i/ & 6 & 3 &  &  &  &  &  & 0 &  &  & $D_4^{(3)}$ \vspace{-0.25cm}\\
   &  &  & 1 & * & 1 & 3 & 0 &  & 3 & 1/2 & \\
\hline 
   &  &  & 0 & * & 1 & 1 & 0 &  & 1 & 1/2 & \vspace{-0.25cm}\\
j/ & 6 & 3 &  &  &  &  &  & 0 &  &  & $G_2^{(1)}$ \vspace{-0.25cm}\\
   &  &  & 1 & * & 3 & 1 & 0 &  & 1/3 & 3/2 & \\
\hline 
   &  &  & 0 & * & 2 & 1 & 0 &  & 1 & 1 & \vspace{-0.25cm}\\
k/ & 6 & 3 &  &  &  &  &  & 0 &  &  & $D_3^{(2)}$ \vspace{-0.25cm}\\
   &  &  & 1 & * & 2 & 1 & 0 &  & 1 & 1 & \\
\hline 
   &  &  & 0 & * & 1 & 2 & 0 &  & 1 & 1/2 & \vspace{-0.25cm}\\
l/ & 6 & 3 &  &  &  &  &  & 0 &  &  & $A_4^{(2)}$ \vspace{-0.25cm}\\
   &  &  & 1 & * & 2 & 1 & 0 &  & 1/4 & 1 & \\
\hline 
   &  &  & 0 & * & 1 & 2 & 0 &  & 1 & 1/2 & \vspace{-0.25cm}\\
m/ & 6 & 3 &  &  &  &  &  & 0 &  &  & $C_2^{(1)}$ \vspace{-0.25cm}\\
   &  &  & 1 & * & 1 & 2 & 0 &  & 1 & 1/2 & \\
\hline 
   &  &  & 0 & * & 1 & 1 & 0 &  & 1 & 1/2 & \vspace{-0.25cm}\\
n/ & 6 & 2 &  &  &  &  &  & -1/2 &  &  &($n_4=0$) $A_3^{(1)}$ \vspace{-0.25cm}\\
   &  &  & 1 & * & 1 & 1 & 0 &  & 1 & 1/2 & \\
\hline 
 &  &  & 0 & * & 1 & 1 & 0 &  & 1 & 1/2 & \\
o/ & 8  & 3 & 1 & * & 1 & 1 & 0 & 0 & 1 & 1/2 & $D_4^{(1)}$ \\
 &  &  & 2 & 0 & 1 & 1 & -1/2 &  & 1/2 & 1 & \\
\hline 
     &  &  & 0 & * & 1 & 2 & 0 &  & 1 & 1/2 &\\
p/ & 8 & 4 & 1 & * & 1 & 1 & 0 & 0 & 1/2 & 1/2 & $A_5^{(2)}$ \\
     &  &  & 2 & * & 1 & 1 & 0 &  & 1/2 & 1/2 & \\
\hline 
      &  &  & 0 & * & 2 & 1 & 0 &  & 1 & 1 & \\
q/ & 8  & 4 & 1 & * & 1 & 1 & 0 & 0 & 2 & 1/2 & $B_3^{(1)}$\\
      &  &  & 2 & * & 1 & 1 & 0 &  & 2 & 1/2 & \\
\hline 
       &  &  & 0 & * & 1 & 1 & 0 &  & 1 & 1/2 & \\
       &  &  & 1 & * & 1 & 1 & 0 &  & 1 & 1/2 & \vspace{-0.25cm}\\
r/ & 10  & 5 &   &   &   &   &   & 0 &  &   & $D_4^{(1)}$ \vspace{-0.25cm}\\
       &  &  & 2 & * & 1 & 1 & 0 &  & 1 & 1/2 & \\
       &  &  & 3 & * & 1 & 1 & 0 &  & 1 & 1/2 & \\
\hline 
\end{tabular}\par}
\begin{center}
{\small \underline{Table 1.2} - Solutions for the even case with restrictions, and the corresponding algebras\\
\vspace{-0.04cm}of rank $r=n-1-R$. $R$ : number of restrictions ($R=s+1$ if $A_{n/2-1} \ne 0$, else $R=s+2$). \\
\vspace{-0.04cm}$*$ stands for one of the $s+1$ restrictions as defined in (\ref{def-rest-even})}
\end{center}

\newpage
\vspace{5mm}

\centerline{\epsfig{file=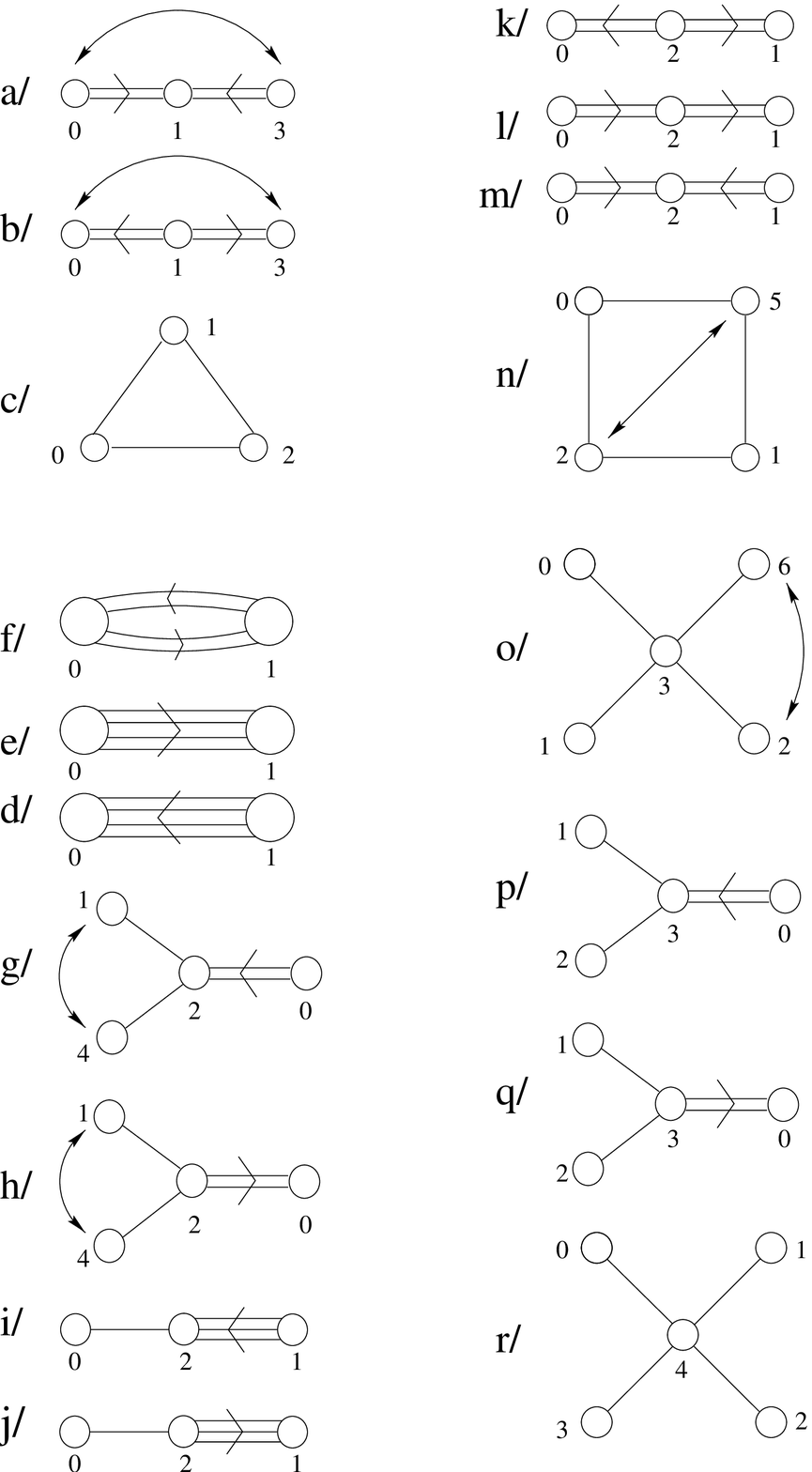,height=190mm,width=95mm}}
\vspace{5mm}
\begin{center}
{\small \underline{Figure 1.2}  - Dynkin diagrams corresponding to affine Lie algebras Table 1.2 .}
\end{center}

\newpage
\vspace{0.3cm}
{\centering \begin{tabular}{|c|c|c|c||c|c|c||c|c|c||c|}
\hline 
& $n$ & $R$ &  a & $n_{1}^{a}$ & $n_{2}^{a}$ & $n_{3}^{a}$ & $A_{a}/B_{0}$ & $B_{a}/B_{0}$ & $\delta _{a}$ & Algebra\\
\hline 
\hline 
a/ & 3 & 1 & 0 & * & 4 & 1 & 0 & 1 & 1 & $A_2^{(2)}$ \\
\hline 
b/ & 3 & 1 & 0 & * & 1 & 4 & 0 & 1 & 1/4 & $A_2^{(2)}$ \\
\hline 
c/ & 3 & 1 & 0 & * & 2 & 2 & 0 & 1 & 1/2 & $A_1^{(1)}$ \\
\hline 
 &  &  & 0 & * & 1 & 2 & 0 & 1 & 1/4 & \vspace{-0.25cm}\\
d/ & 5 & 1 &  &  &  &  &  &  &  & $A_5^{(2)}$ \vspace{-0.25cm}\\
 &  &  & 1 & 0 & 1 & 1 & -1/4 & 1/4 & 1/2 & \\
\hline 
   &  &  & 0 & * & 2 & 1 & 0 & 1 & 1/2 & \vspace{-0.25cm}\\
e/ & 5 & 1 &  &  &  &  &  &  &  & $B_3^{(1)}$ \vspace{-0.25cm}\\
   &  &  & 1 & 0 & 1 & 1 & -1 & 1 & 1/2 & \\
\hline 
   &  &  & 0 & * & 1 & 1 & 0 & 1 & 1/4 & \vspace{-0.25cm}\\
f/ & 5 & 2 &  &  &  &  &  &  &  & $D_4^{(3)}$ \vspace{-0.25cm}\\
   &  &  & 1 & * & 1 & 3 & 0 & 3 & 1/4 & \\
\hline 
   &  &  & 0 & * & 1 & 1 & 0 & 1 & 1/4 & \vspace{-0.25cm}\\
g/ & 5 & 2 &  &  &  &  &  &  &  & $G_2^{(1)}$ \vspace{-0.25cm}\\
   &  &  & 1 & * & 3 & 1 & 0 & 1/3 & 3/4 & \\
\hline 
   &  &  & 0 & * & 1 & 2 & 0 & 1 & 1/4 & \vspace{-0.25cm}\\
h/ & 5 & 2 &  &  &  &  &  &  &  & $C_2^{(1)}$ \vspace{-0.25cm}\\
   &  &  & 1 & * & 1 & 2 & 0 & 1 & 1/4 & \\
\hline 
   &  &  & 0 & * & 1 & 2 & 0 & 1 & 1/4 & \vspace{-0.25cm}\\
i/ & 5 & 2 &  &  &  &  &  &  &  & $A_4^{(2)}$ \vspace{-0.25cm}\\
   &  &  & 1 & * & 2 & 1 & 0 & 1/4 & 1/2 & \\
\hline 
   &  &  & 0 & * & 2 & 1 & 0 & 1 & 1/2 & \vspace{-0.25cm}\\
j/ & 5 & 2 &  &  &  &  &  &  &  & $D_3^{(2)}$ \vspace{-0.25cm}\\
   &  &  & 1 & * & 2 & 1 & 0 & 1 & 1/2 & \\
\hline 
 &  &  & 0 & * & 1 & 1 & 0 & 1 & 1/4 & \\
k/ & 7  & 2 & 1 & * & 1 & 1 & 0 & 1 & 1/4 & $D_4^{(1)}$ \\
 &  &  & 2 & 0 & 1 & 1 & -1/2 & 1/2 & 1/2 & \\
\hline 
 &  &  & 0 & * & 1 & 2 & 0 & 1 & 1/4 &\\
l/ & 7 & 3 & 1 & * & 1 & 1 & 0 & 1/2 & 1/4 & $A_5^{(2)}$ \\
 &  &  & 2 & * & 1 & 1 & 0 & 1/2 & 1/4 & \\
\hline 
 &  &  & 0 & * & 2 & 1 & 0 & 1 & 1/2 & \\
m/ & 7  & 3 & 1 & * & 1 & 1 & 0 & 2 & 1/4 & $B_3^{(1)}$\\
 &  &  & 2 & * & 1 & 1 & 0 & 2 & 1/4 & \\
\hline 
 &  &  & 0 & * & 1 & 1 & 0 & 1 & 1/4 & \\
 &  &  & 1 & * & 1 & 1 & 0 & 1 & 1/4 & \vspace{-0.25cm}\\
n/ & 9  & 4 &  &  &  &  &  &  &  & $D_4^{(1)}$ \vspace{-0.25cm}\\
 &  &  & 2 & * & 1 & 1 & 0 & 1 & 1/4 & \\
 &  &  & 3 & * & 1 & 1 & 0 & 1 & 1/4 & \\
\hline 
\end{tabular}\par}
\begin{center}
{\small \underline{Table 1.3} - Solutions for the odd case with restrictions, and the corresponding algebras.\\
\vspace{-0.04cm}of rank $r=n-1-R$. $R$ : number of restrictions : $R=s+1$. \\
\vspace{-0.04cm}$*$ stands for one of the $s+1$ restrictions as defined in (\ref{def-rest-odd}).}
\end{center}
\vspace{1cm}
\newpage
\vspace{5mm}

\centerline{\epsfig{file=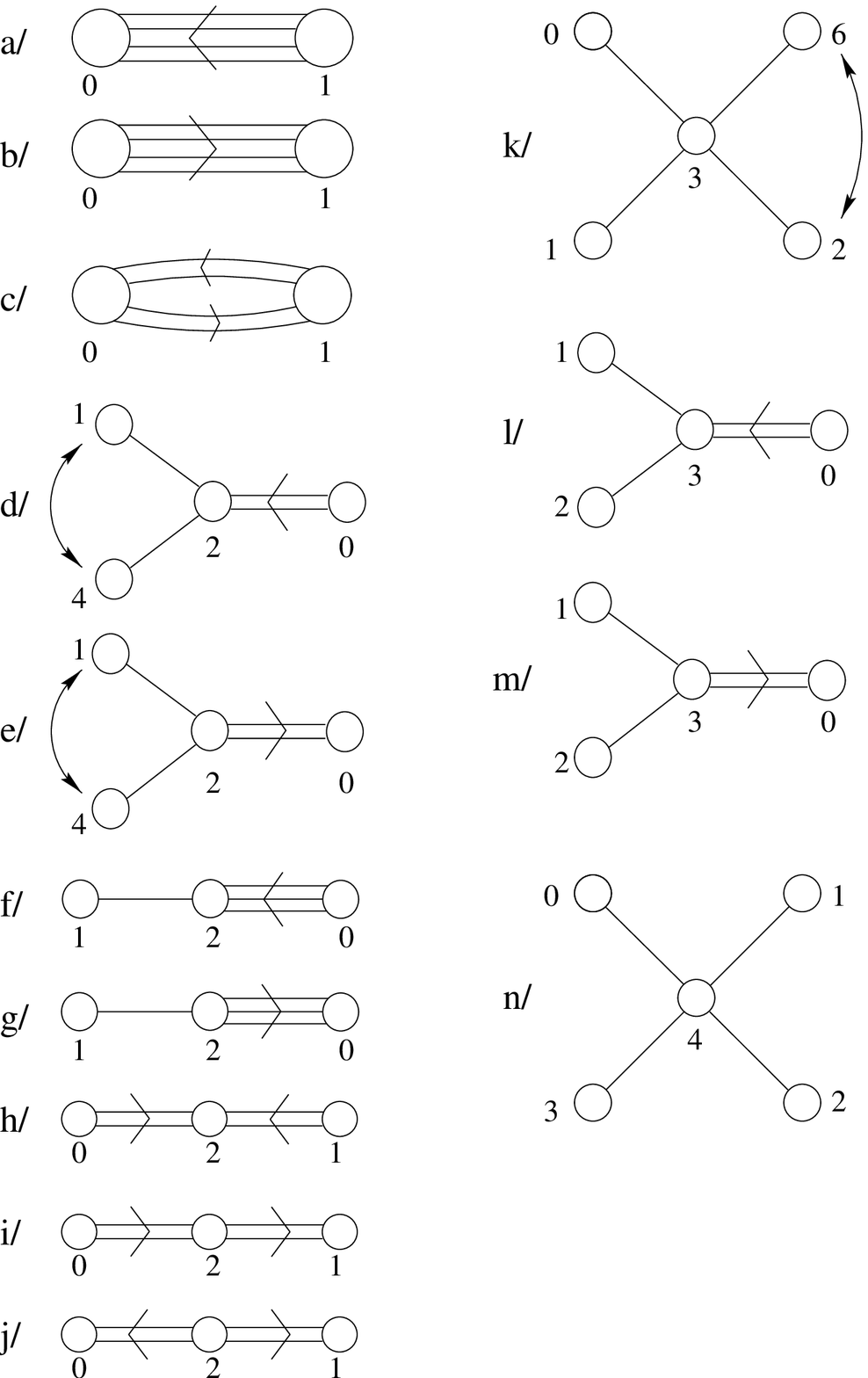,height=160mm,width=95mm}}
\vspace{5mm}
\begin{center}
{\small \underline{Figure 1.3}  - Dynkin diagrams corresponding to affine Lie algebras Table 1.3 .}
\end{center}

\newpage
\centerline{\bf \large Appendix B}
\vspace{0.3cm}
\underline{\bf $\bullet \ \  n$ even.} \\
The set of ``parametrized'' roots for $n$ even reads, with $a \in \{ 0, \dots , \frac{n}{2}-2  \}$ :
\beqa
&&{\bf r}_a = \big[0,...,0,i\frac{\alpha_a}{\beta_0},\frac{\beta_a}{\beta_0},0,...,0\big],\ \ \ \ {\bf r}_{a+\frac{n}{2}}=\big[0,...,0,-i\frac{\alpha_a}{\beta_0},\frac{\beta_a}{\beta_0},0,...,0\big],\label{roots} \nonumber \\
&&{\bf r}_{\frac{n}{2}-1}= \big[0,-\frac{m_0}{m_{\frac{n}{2}-1}},...,0,-\frac{\beta_a m_a}{\beta_0m_{\frac{n}{2}-1}},...,0,-\frac{\beta_{\frac{n}{2}-2} m_{\frac{n}{2}-2}}{\beta_0m_{\frac{n}{2}-1}},i\frac{\alpha_{\frac{n}{2}-1}}{\beta_0}\big] , \\
&&{\bf r}_{n-1}= \big[0,-\frac{m_0}{m_{\frac{n}{2}-1}},...,0,-\frac{\beta_a m_a}{\beta_0m_{\frac{n}{2}-1}},...,0,-\frac{\beta_{\frac{n}{2}-2} m_{\frac{n}{2}-2}}{\beta_0m_{\frac{n}{2}-1}},-i\frac{\alpha_{\frac{n}{2}-1}}{\beta_0}\big].\nonumber
\eeqa
 For ${\cal P}$, the set of  dual ``parametrized'' co-weights defined by (\ref{system-co-root}) reads, with $ a \in \{ 0, \dots, \frac{n}{2}-2 \} $ :
\bea
\label{coroot-C-even}
\beta_0^{\vee} {\bf \omega}_{a;\frac{n}{2}-1}^{\vee} &=& [ 0, \dots , 0,  \frac{1}{2i\alpha_a}, \frac{1}{2\beta_a} , 0 , \dots, 0, -\frac{m_a}{2im_{\frac{n}{2}-1} \alpha_{\frac{n}{2}-1}} ], \nonumber\\
\beta_0^{\vee} {\bf \omega}_{a+\frac{n}{2};\frac{n}{2}-1}^{\vee} &=&  [ 0, \dots , 0, -\frac{1}{2i\alpha_a}, \frac{1}{2\beta_a} , 0 , \dots, 0 , -\frac{m_a}{2im_{\frac{n}{2}-1} \alpha_{\frac{n}{2}-1}} ], \\
\beta_0^{\vee} {\bf \omega}_{n-1;\frac{n}{2}-1}^{\vee} &=& [ 0, \dots \dots  , 0,- \frac{1}{i\alpha_{\frac{n}{2}-1}} ]. \nonumber
\eea
For $\overline{\cal P}$, the set of dual ``parametrized'' co-weights reads, with $a \in \{ 1, \dots, \frac{n}{2}-2\} $ is  :
\bea
\label{co-root-Cbar-even}
\beta_0^{\vee} {\bf \omega}_{a;0}^{\vee} &=& [-\frac{m_a}{2im_0\alpha_0} , -\frac{m_a}{2m_0\beta_0} , 0, \dots ,0, \frac{1}{2i\alpha_a}, \frac{1}{2\beta_a }, 0 \dots,0 ],  \nonumber \\
\beta_0^{\vee} {\bf \omega}_{a+\frac{n}{2};0}^{\vee} &=& [-\frac{m_a}{2im_0\alpha_0} , -\frac{m_a}{2m_0\beta_0} , 0, \dots ,0, -\frac{1}{2i\alpha_a}, \frac{1}{2\beta_a }, 0 \dots,0 ],  \nonumber \\
\beta_0^{\vee} {\bf \omega}_{\frac{n}{2}-1;0}^{\vee} &=& [ -\frac{m_{\frac{n}{2}-1}}{2im_0\alpha_0} ,  -\frac{m_{\frac{n}{2}-1}}{2m_0\beta_0} , 0 , \dots , 0, \frac{1}{2i\alpha_{\frac{n}{2}-1}} ], \\
\beta_0^{\vee} {\bf \omega}_{n-1;0}^{\vee} &=& [ -\frac{m_{\frac{n}{2}-1}}{2im_0 \alpha_0} ,  -\frac{m_{\frac{n}{2}-1}}{2m_0\beta_0} , 0 , \dots , 0, -\frac{1}{2i\alpha_{\frac{n}{2}-1}} ], \nonumber  \\
\beta_0^{\vee} {\bf \omega}_{\frac{n}{2};0}^{\vee} &=& [ -\frac{1}{i\alpha_0}, 0 \dots , 0 ]. \nonumber 
\eea

Similarly, ${\bf \omega}_a$ are obtained using substitutions 
$\alpha_a \to \alpha^{\vee}_a$, $\beta_a \to \beta^{\vee}_a$ and $m_a \to m^{\vee}_a$.

\underline{\bf $\bullet \ \ n$ odd.}\\
For $n$ odd the set of ``parametrized'' roots is given by ( with $a \in \{ 0,\dots, \frac{n-1}{2}-1 \} $) :
\bea
&&{\bf r}_a = \big[0,...,0,i\frac{\alpha_a}{\beta_0},\frac{\beta_a}{\beta_0},0,...,0\big],\ \ \ \ {\bf r}_{a+\frac{n+1}{2}}=\big[0,...,0,-i\frac{\alpha_a}{\beta_0},\frac{\beta_a}{\beta_0},0,...,0\big],\label{roots-odd} \nonumber \\
&&{\bf r}_{\frac{n-1}{2}}= \big[0,-\frac{2m_0}{m_{\frac{n}{2}-1}},...,0,-\frac{2\beta_a m_a}{\beta_0m_{\frac{n}{2}-1}},...,0,-\frac{2\beta_{\frac{n}{2}-2} m_{\frac{n}{2}-2}}{\beta_0m_{\frac{n}{2}-1}} \big].
\eea
For ${\cal P}$, the set of dual ``parametrized'' co-weights is then, with $a \in \{0, \dots , \frac{n-1}{2}-1 \}$ :
\bea
\label{coroot-C-odd}
\beta_0^{\vee} {\bf \omega}_{a;\frac{n-1}{2}}^{\vee} &=& [ 0 , \dots , 0 ,\frac{1}{2i \alpha_a} , \frac{1}{2\beta_a} , 0 , \dots , 0 ], \nonumber \\
\beta_0^{\vee} {\bf \omega}_{a+\frac{n+1}{2};\frac{n-1}{2}}^{\vee} &=& [ 0 , \dots , 0 ,-\frac{1}{2i \alpha_a} , \frac{1}{2\beta_a} , 0 , \dots , 0 ].
\eea
For $\overline{\cal P}$, the set of dual ``parametrized'' co-weights is, with $ a\in \{1, \dots,\frac{n-1}{2}-1\} $ :
\bea
\label{co-root-Cbar-odd}
\beta_0^{\vee} {\bf \omega}_{a;0}^{\vee} &=& [-\frac{m_a}{2im_0 \alpha_0} ,-\frac{m_a}{2m_0 \beta_0} , 0, \dots ,0, \frac{1}{2i\alpha_a}, \frac{1}{2\beta_a }, 0 \dots,0 ],  \nonumber \\
\beta_0^{\vee} {\bf \omega}_{a+\frac{n+1}{2};0}^{\vee} &=& [-\frac{m_a}{2im_0 \alpha_0} ,-\frac{m_a}{2m_0 \beta_0} , 0, \dots ,0, -\frac{1}{2i\alpha_a}, \frac{1}{2\beta_a }, 0 \dots,0 ], \\
\beta_0^{\vee} {\bf \omega}_{\frac{n-1}{2};0}^{\vee} &=& [ -\frac{m_{\frac{n-1}{2}}}{2im_0 \alpha_0}, -\frac{m_{\frac{n-1}{2}}}{2 m_0 \beta_0} , 0,  \dots , 0 ], \nonumber  \\
\beta_0^{\vee} {\bf \omega}_{\frac{n-1}{2}+1;0}^{\vee} &=& [ -\frac{1}{i\alpha_0}, 0 \dots , 0 ]. \nonumber 
\eea

\end{document}